\documentclass[RNAAS]{aastex63}

\usepackage[hang,flushmargin,stable]{footmisc} 
\usepackage{graphicx}
\usepackage{txfonts}
\usepackage{xcolor}
\newcommand{\orcid}[1]{\href{https://orcid.org/#1}{\includegraphics[width=8pt]{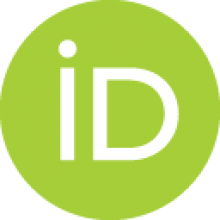}}}

\shortauthors{Ye Feng et al.}
\graphicspath{{./}{figures/}}
\begin{document}

\title{Estimating the spin of the black hole candidate MAXI J1659-152 with the X-ray continuum-fitting method}

\author{Ye Feng\orcid{0000-0002-6961-8082}}
\affiliation{Key Laboratory for Computational Astrophysics, National Astronomical Observatories, Chinese Academy of Sciences, Datun Road A20, Beijing 100012, China; \href{mailto:yefeng@nao.cas.cn}{yefeng@nao.cas.cn}, \href{mailto:lgou@nao.cas.cn}{lgou@nao.cas.cn}}
\affiliation{School of Astronomy and Space Sciences, University of Chinese Academy of Sciences, Datun Road A20, Beijing 100049, China}

\author{Xueshan Zhao\orcid{0000-0002-3281-5306}}
\affiliation{Key Laboratory for Computational Astrophysics, National Astronomical Observatories, Chinese Academy of Sciences, Datun Road A20, Beijing 100012, China; \href{mailto:yefeng@nao.cas.cn}{yefeng@nao.cas.cn}, \href{mailto:lgou@nao.cas.cn}{lgou@nao.cas.cn}}
\affiliation{School of Astronomy and Space Sciences, University of Chinese Academy of Sciences, Datun Road A20, Beijing 100049, China}
 
\author{Lijun Gou\orcid{0000-0003-3057-5860}}
\affiliation{Key Laboratory for Computational Astrophysics, National Astronomical Observatories, Chinese Academy of Sciences, Datun Road A20, Beijing 100012, China; \href{mailto:yefeng@nao.cas.cn}{yefeng@nao.cas.cn}, \href{mailto:lgou@nao.cas.cn}{lgou@nao.cas.cn}}
\affiliation{School of Astronomy and Space Sciences, University of Chinese Academy of Sciences, Datun Road A20, Beijing 100049, China}

\author{Jianfeng Wu\orcid{0000-0001-7349-4695}}
\affiliation{Department of Astronomy, Xiamen University, Xiamen, Fujian 361005, China}

\author{James F. Steiner\orcid{0000-0002-5872-6061}}
\affiliation{Harvard-Smithsonian Center for Astrophysics, 60 Garden Street, Cambridge, MA 02138, USA}

\author{Yufeng Li\orcid{0000-0002-4135-074X}}
\affiliation{Key Laboratory for Computational Astrophysics, National Astronomical Observatories, Chinese Academy of Sciences, Datun Road A20, Beijing 100012, China; \href{mailto:yefeng@nao.cas.cn}{yefeng@nao.cas.cn}, \href{mailto:lgou@nao.cas.cn}{lgou@nao.cas.cn}}
\affiliation{School of Astronomy and Space Sciences, University of Chinese Academy of Sciences, Datun Road A20, Beijing 100049, China}

\author{Zhenxuan Liao\orcid{0000-0003-2221-0490}}
\affiliation{Key Laboratory for Computational Astrophysics, National Astronomical Observatories, Chinese Academy of Sciences, Datun Road A20, Beijing 100012, China; \href{mailto:yefeng@nao.cas.cn}{yefeng@nao.cas.cn}, \href{mailto:lgou@nao.cas.cn}{lgou@nao.cas.cn}}
\affiliation{School of Astronomy and Space Sciences, University of Chinese Academy of Sciences, Datun Road A20, Beijing 100049, China}

\author{Nan Jia\orcid{0000-0002-5487-880X}}
\affiliation{Key Laboratory for Computational Astrophysics, National Astronomical Observatories, Chinese Academy of Sciences, Datun Road A20, Beijing 100012, China; \href{mailto:yefeng@nao.cas.cn}{yefeng@nao.cas.cn}, \href{mailto:lgou@nao.cas.cn}{lgou@nao.cas.cn}}
\affiliation{School of Astronomy and Space Sciences, University of Chinese Academy of Sciences, Datun Road A20, Beijing 100049, China}

\author{Yuan Wang\orcid{0000-0001-9604-0370}}
\affiliation{Key Laboratory for Computational Astrophysics, National Astronomical Observatories, Chinese Academy of Sciences, Datun Road A20, Beijing 100012, China; \href{mailto:yefeng@nao.cas.cn}{yefeng@nao.cas.cn}, \href{mailto:lgou@nao.cas.cn}{lgou@nao.cas.cn}}
\affiliation{School of Astronomy and Space Sciences, University of Chinese Academy of Sciences, Datun Road A20, Beijing 100049, China}

\begin{abstract}
As a transient X-ray binary, MAXI J1659-152 contains a black hole candidate as its compact star. MAXI J1659-152 was discovered on 2010 September 25 during its only known outburst. Previously-published studies of this outburst indicate that MAXI J1659-152 may have an extreme retrograde spin, which, if confirmed, would provide an important clue as to the origin of black hole spin. In this paper, utilizing updated dynamical binary-system parameters (i.e. the black hole mass, the orbital inclination and the source distance) provided by \cite{Torres2021}, we analyze 65 spectra of MAXI J1659-152 from \emph{RXTE}/PCA, in order to assess the spin parameter.  With a final selection of 9 spectra matching our $f_{\mathrm{sc}} \lesssim 25 \%$, soft-state criteria, we apply a relativistic thin disk spectroscopic model \texttt{kerrbb2} over 3.0-45.0 keV. We find that inclination angle correlates inversely with spin, and, considering the possible values for inclination angle, we constrain spin to be $-1 < a_{*} \lesssim 0.44$ at 90\% confidence interval via X-ray continuum-fitting. We can only rule out an extreme prograde (positive) spin. We confirm that an extreme retrograde solution is possible and is not ruled out by considering accretion torques given the young age of the system.

\end{abstract}


\section{1. INTRODUCTION}
\label{section:1}
Spin is one of the most important basic physical quantities that characterizes a black hole. Knowledge of this parameter is essential for understanding the physics governing black holes and their phenomena such as jets (\citealt{Narayan2012}). Spin is described by a dimensionless parameter $a_{*} = cJ/GM^2$, which lies in a range from -1 to 1 (where $c$ is the speed of light, $J$ is the angular momentum of the black hole, $G$ is the gravitational constant, and $M$ is the black hole mass). 

The spin of a black hole X-ray binary (BHXRB) is generally measured in one of two spectroscopic methods.  The first technique, X-ray ``reflection fitting'' was pioneered by \citet{Fabian1989}. Subsequently, the X-ray ``continuum-fitting'' method was proposed by \citet{Zhang1997}. In recent years, non spectroscopic techniques for measuring spin have also been proposed and explored. For instance, the high-frequency quasi-periodic oscillations (HFQPO) method (\citealt{Wagoner2001}, \citealt{Motta2014}), X-ray polarization method (\citealt{Dov2008}), and methods for other black holes not in X-ray binaries such as the black hole horizon method (\citealt{Dokuchaev2019}), and numerous gravitational waveform measurements of black hole mergers (See Gravitational Wave Open Science Center\footnote{\url{https://www.gw-openscience.org/catalog/GWTC-1-confident/html/}}).
	
Of these methods, only X-ray continuum-fitting (hereafter CF)  and X-ray reflection fitting (also known as the $\text {Fe K}  \alpha$  method) are widely used in BHXRBs. To date, less than 100 BHXRBs have been found (See BlackCAT\footnote{\url{http://www.astro.puc.cl/BlackCAT/}} for details), the spin of more than two dozen BHXRBs (including persistent and transient sources) in and around the Galaxy have been measured by at least one method. For example, GRS 1915+105 (\citealt{McClintock2006}), M33 X-7 (\citealt{Liu2008}), LMC X-1 (\citealt{Gou2009}), A 0620-00 (\citealt{Gou2010}), XTE J1550-564 (\citealt{Steiner2011}), Cyg X-1 (\citealt{Gou2014}), GS 1124-683 (\citealt{Chen2016}), XTE J1752-223 (\citealt{Javier2018}), 4U 1543-47 (\citealt{Dong2020}) and so on.

The CF method is based on the classic relativistic thin disk model (\citealt{Novikov1973}). Through fitting the thermal continuum emission of the accretion disk, the inner-disk radius is constrained. By adopting the usual assumption that the inner radius of the accretion disk extends to the innermost stable circular orbit (ISCO), we can ascertain the spin based on the unique mapping between ISCO and spin as revealed in \cite{Bardeen1972}. The CF method relies on accurate measurement of the system parameters (such as black hole mass ($M$), the inclination of the accretion disk ($i$, which is assumed to be equal to the orbital inclination) and the source distance ($D$)). The $\text {Fe  K} \alpha$ fitting method also measures the inner-disk radius by modeling the relativistic reflection spectrum and is only weakly dependent upon inclination, which may also be fitted alongside spin. The CF method is generally only applicable to stellar-mass black holes, while $\text {Fe K}  \alpha$ fitting is widely used for both stellar-mass black holes and supermassive black holes. Aside from mutual reliance on the association between inner-disk radius and ISCO, both methods are independent and can be used to cross-check one another. In this paper, we apply the CF method to estimate the spin of MAXI J1659-152.

As the capability of spin measurement by the CF method relies upon ensuring that the inner radius of the accretion disk extends to the ISCO, we implement a screening to identify data for which that requirement is most sound: namely, high/soft (HS) state spectra whose emission is dominated by the accretion disk (e.g., \citealt{McClintock2006}). This also avoids potential contamination by a strong Compton component.  We additionally require that the disk is geometrically thin by requiring that the dimensionless luminosity $l=L(a_{*}, \dot{M})/ L_{\mathrm{Edd}}<$0.3 (where $H$ is the thickness of the disk, $R$ is the local disk radius and $l$ is the bolometric Eddington-scaled luminosity; e.g., \citealt{Gou2011}). To ensure the data are sufficiently dominated by the thermal disk emission, we apply a bound on the relative strength of the nonthermal emission   (\citealt{Steiner2009}; \citealt{Steiner2009b}) which has been applied in similar analysis of other black-hole systems (e.g., \citealt{Steiner2011}; \citealt{Chen2016}; \citealt{Zhao2020}). Briefly the proportion of thermal photons which scatter in the corona (the ``scattering-fraction'') $f_{\mathrm{sc}} \lesssim 25 \%$, where $f_{\mathrm{sc}}$ is a parameter of the Comptonization model \texttt{simpl} in \texttt{XSPEC}.  

Generally, BHXRBs can be divided into high-mass X-ray binaries (HMXBs) and low-mass X-ray binaries (LMXBs) according to the mass of the black hole's companion star. Typically, HMXBs are fueled by the capture of strong winds from the companion star (\citealt{Shakura2015}). However, LMXBs are usually fueled via Roche-lobe overflow (RLO) in a stream through the first Lagrangian point (L1) (\citealt{Savonije1978}). In theory, natal kicks may lead to a random initial distribution of prograde/retrograde black hole X-ray binaries. However, the strong majority of black holes observed have a positive spin. For these black holes with positive spin, \cite {Nielsen2016} shows that black holes in HMXBs usually have higher spin, while LMXB black hole spins range from very low to very high.  Notably, the binary black holes (BBHs) detected by LIGO/Virgo by O3a appear to exhibit spin magnitudes close to zero (see \citealt{Abbott2020} for details on BBHs). At present, there is scant information on retrograde-spin systems. There are four black hole systems for whom a retrograde spin has been suggested as likely: IGR J17091-3624 (\citealt{Rao2012}), Swift J1910.2-0546 (\citealt{Reis2013}), GS 1124-683 (\citealt{Morningstar2014}), XMMU J004243.6+412519 (\citealt{Middleton2014}). Amidst these, a revised estimate of GS 1124-683 has offered instead a moderate prograde spin (\citealt{Chen2016}).

\setlength{\tabcolsep}{2.5mm}
\begin{deluxetable*}{ccccccc}
\tablecaption{Properties of SP1-SP9\label{table:1}}
\tablewidth{700pt}
\centering
\tabletypesize{\scriptsize}
\tablehead{
\colhead{Spec.} & \colhead{ObsID} & \colhead{MJD} & \colhead{Start Time} & 
\colhead{End Time} & \colhead{Exposure}  & \colhead{Count rates}
\\
\cline{4-5}
\colhead{} & \colhead{} & 
\colhead{} & \colhead{} & 
\colhead{} & \colhead{(s)}  & \colhead{(cts~$\text{s}^{-1}$)}
} 
\startdata
SP1	&	95118-01-03-00	&	55486 	&	2010-10-17~19:01:04	&	2010-10-17~19:42:40	&	2345 	&      248.8 \\
SP2	&	95118-01-05-00	&	55488 	&	2010-10-19~00:27:44	&	2010-10-19~01:17:52	&	2696	&	249.9 \\
SP3	&	95118-01-05-01	&	55488 	&	2010-10-19~20:52:32	&	2010-10-19~21:34:56	&	2249 	&	222.3 \\
SP4	&	95118-01-06-00	&	55489 	&	2010-10-20~06:20:48	&	2010-10-20~07:23:44	&	2938 	&	221.0 \\
SP5	&	95118-01-07-01	&	55490 	&	2010-10-21~02:45:04	&	2010-10-21~03:33:04	&	2764 	&	247.9 \\
SP6	&	95118-01-13-00	&	55496 	&	2010-10-27~12:26:24	&	2010-10-27~13:07:44	&	2139 	&	158.1 \\
SP7	&	95118-01-14-00	&	55497 	&	2010-10-28~12:03:12	&	2010-10-28~12:16:48	&	749 	&	140.3 \\
SP8	&	95118-01-15-00	&	55498 	&	2010-10-29~11:40:48	&	2010-10-29~12:00:48	&	1126 	&	140.3 \\
SP9	&	95118-01-15-01	&	55499 	&	2010-10-30~07:57:36	&	2010-10-30~08:21:36	&	1378 	&      137.0 \\
\enddata
{Notes.\\
In columns 2-7, we show undermentioned information: the observation ID (ObsID), Modified Julian Date (MJD), the start time, the end time, the exposure time in units of s, and the net count rates of SP1-SP9 for PCU2 top layer measured at 3.0-45.0 keV in units of cts~$\text{s}^{-1}$. 
}
\end{deluxetable*}

MAXI J1659-152 (hereafter MAXI J1659) is a Galactic black hole candidate X-ray binary (\citealt{Postigo2010}). As a transient source, MAXI J1659 has been dormant during the monitoring of the Gas Slit Camera of the \emph{Monitor of All-sky X-ray Image} (\emph{MAXI}/GSC) over the past 11 years. On September 25th, 2010, it was first detected  entering a seven-month outburst (\citealt{Homan2013}). The discovery was reported by the Burst Alert Telescope of the \emph{Neil Gehrels Swift Observatory} (\emph{Swift}/BAT) (\citealt{Mangano2010}) and \emph{MAXI}/GSC (\citealt{Negoro2010}) in the gamma-ray and X-ray bands, individually. Later, optical (\citealt{Postigo2010}), radio (\citealt{Plotkin2013}) and near-infrared (\citealt{Kaur2012}) data were also obtained. Its fast timing behaviour flagged it as a likely stellar-mass black hole candidate (\citealt{Yamaoka2012}). \cite {Kuulkers2013} report that the orbital period of the binary system is just 2.42 hours and that the binary system consists of an M5 dwarf companion with the mass of 0.15-0.25$~\mathrm{M}_{\odot}$ and a radius of 0.2-0.25$~\mathrm{R}_{\odot}$, establishing MAXI J1659 as a low-mass X-ray binary. Since its outburst, the black hole has attracted much attention, and there has been extensive discussion about its mass (\citealt{Kennea2011}; \citealt{Yamaoka2012}; \citealt{Shaposhnikov2011}; \citealt{Rao2015}; \citealt{Molla2016}), inclination (\citealt{Kennea2011}; \citealt{Yamaoka2012}) and distance (\citealt{Kennea2011}; \citealt{Kaur2012}; \citealt{Kong2012}; \citealt{Jonker2012}; \citealt{Kuulkers2013}). Whereas, due to the faintness of the quiescent optical counterpart and short orbital period of MAXI J1659, it is difficult to measure its dynamic parameters (\citealt{Torres2021}). Recently, \cite{Rout2020} claims that the black hole's spin is retrograde and that it may be maximal. Through the $H_{\alpha}$ emission, \cite {Torres2021} find a radial velocity semi-amplitude of the donor of $K_{2}=750\pm 80 \text{ km} \text{ s}^{-1}$. Furthermore, \cite {Torres2021} obtain $q=M_{2}/M_{1}=0.02-0.07$ based on the interdependence between the ratio of line double-peak line separation (DP) and FWHM with $q$. Since the system lacks eclipses, based on modeling the light curve, \cite {Torres2021} further consider a case where the disk's outer rim occults the central X-ray source from the donor, and compares the $H_{\alpha}$ line profile with that of multiple black hole systems. Taken together, the inclination is restricted in the range $70^{\circ} \lesssim i \lesssim  80^{\circ}$. In addition, the detection of X-ray absorption dips during the early outburst indicates a high inclination (e.g., \citealt{Kuulkers2013}). The mass of MAXI J1659 is constrained between $5.7 \pm 1.8 ~\mathrm{M}_{\odot}$ ($i=70^{\circ}$) and $4.9 \pm 1.6 ~\mathrm{M}_{\odot}$ ($i=80^{\circ}$) at the confidence level of $68.3\%$ (for details, see \citealt{Torres2021}). In recent years, measuring properties of the $H_{\alpha}$ emission line profile emitted by the quiescent accretion disk to estimate $K_{2}$ and $q$ has become more mature and reliable (\citealt{Casares2015}, \citeyear{Casares2016}, \citeyear{Casares2018}; \citealt{Casares2018b}), which also inspires us to revisit the spin of MAXI J1659.

Our work is based upon the CF method, and utilizes the latest measurement results of \cite {Torres2021}, adopting two limiting solutions: $M=5.7 \pm 1.8 ~\mathrm{M}_{\odot}$ (1$\sigma$), $i=70.0^{\circ}$, $D=6 \pm 2$~kpc and $M=4.9 \pm 1.6 ~\mathrm{M}_{\odot}$ (1$\sigma$), $i=80.0^{\circ}$, $D=6 \pm 2$~kpc. For the above two sets of system parameters ($M$, $i$, $D$), we fit 9 selected spectra of \emph{RXTE}/PCA of MAXI J1659 throughout its 2010 outburst in the soft state with a relativistic thin disk model \texttt{kerrbb2} (\citealt{McClintock2006}) and constrain its spin. 

The paper is organized as follows. In Section \ref{section:2}, we introduce the data selection and reduction, including both \emph{RXTE}/PCA and \emph{XMM-Newton}/EPIC-pn data. In Section \ref{section:3}, we describe the spectral analysis and results in detail. A discussion is presented in Section \ref{section:4}. We offer our conclusions in Section \ref{section:5}.

\section{2. DATA SELECTION AND REDUCTION}
\label{section:2}

 During MAXI J1659's 2010 outburst, the \emph{Rossi X-ray Timing Explorer} (\emph{RXTE}) carried out a total of 65 continuous observations, which were stored in three program IDs (95358, 95108, and 95118, respectively). We employ the Proportional Counter Array data of \emph{RXTE} (\emph{RXTE}/PCA) (\citealt{Jahoda1996}). In addition, we also consider data from the European Photon Imaging Camera on \emph{X-ray Multi-Mirror Newton} (\emph{XMM-Newton}/EPIC-pn) (\citealt{Str2001}). All data are grouped to achieve a signal-to-noise ratio (S/N) of 25 per energy bin. Spectra are fitted using \texttt{XSPEC v12.11.1}\footnote{\url{https://heasarc.gsfc.nasa.gov/xanadu/xspec/}} (\citealt{Arnaud1996}) using $\chi^2$ statistics. We also check 9 spectra separately using the ``\texttt{ignore bad}'' command in \texttt{XSPEC}, and no bad channels need to be ignored.

\setlength{\tabcolsep}{3mm}
\begin{deluxetable*}{clrrrrlc}
\centering
\tablecaption{The best-fitting parameters for SP1-SP9 with \texttt{crabcor*TBabs*(simpl*diskbb)}}\label{table:2}
\tablewidth{700pt}
\tabletypesize{\scriptsize}
\tablehead{
\colhead{Spec.} & \colhead{ObsID} & \multicolumn{2}{c}{simpl} & \multicolumn{2}{c}{diskbb} & \colhead{$\chi^2_\nu$} & \colhead{$\chi^2$(d.o.f.)}\\
\cline{3-6}
\colhead{} & \colhead{} &\colhead{$\Gamma$} & \colhead{$f_{\rm sc}$} & \colhead{$T_{\rm in}$} & \colhead{$\rm norm$~$(\times 10^{3})$} 
} 
\startdata
SP1	&	95118-01-03-00	&	2.42 	$\pm$	0.02 	&  0.202 	$\pm$	0.003 	&	0.785 	$\pm$	0.006 &	1.26 	$\pm$	0.05 	&	1.296 & 96.48 (68) \\
SP2	&	95118-01-05-00	&	2.42 	$\pm$	0.02 	&  0.240 	$\pm$	0.003 	&	0.771 	$\pm$	0.006 &	1.31 	$\pm$	0.05 	&	1.502 & 115.02 (68)\\
SP3	&	95118-01-05-01	&	2.35 	$\pm$	0.02 	&  0.204 	$\pm$	0.003 	&	0.775 	$\pm$	0.006 &	1.20 	$\pm$	0.05 	&	1.169 & 91.79 (68)\\
SP4	&	95118-01-06-00	&	2.39 	$\pm$	0.02 	&  0.178 	$\pm$	0.003 	&	0.780 	$\pm$	0.005 &	1.22 	$\pm$	0.04 	&	1.094 & 78.80 (68) \\
SP5	&	95118-01-07-01	&	2.39 	$\pm$	0.02 	&  0.259 	$\pm$	0.003 	&	0.781 	$\pm$	0.006 &	1.16 	$\pm$	0.05 	&	1.260 & 93.76 (68) \\
SP6	&	95118-01-13-00	&	2.32 	$\pm$	0.02 	&  0.248 	$\pm$	0.004 	&	0.727 	$\pm$	0.008 &	1.13 	$\pm$	0.07 	&	0.966 & 64.33 (68)\\
SP7	&	95118-01-14-00	&	2.27 	$\pm$	0.05 	&  0.162 	$\pm$	0.005 	&	0.741 	$\pm$	0.011 &	1.08 	$\pm$	0.09 	&	0.964 & 60.46 (68) \\
SP8	&	95118-01-15-00	&	2.36 	$\pm$	0.03 	&  0.214 	$\pm$	0.005 	&	0.697 	$\pm$	0.010 &	1.40 	$\pm$	0.12 &	1.246 & 83.04 (68) \\
SP9	&	95118-01-15-01	&	2.36 	$\pm$	0.03 	&  0.225 	$\pm$	0.004 	&	0.684 	$\pm$	0.010 &	1.49 	$\pm$	0.13 &       0.791 & 54.26 (68) \\
\enddata
{Notes.\\
In columns 2-8, we show summary information: the observation ID (ObsID), the dimensionless photon index of power-law ($\Gamma$), the scattered fraction ($f_{\rm sc}$), the temperature of inner disk radius ($T_{\rm in}$) in the units of keV, the correction factor between the apparent inner disk radius and the realistic radius ($\rm norm$) (see e.g., \citealt{Kubota1998}), the reduced chi-square ($\chi_{\nu}^{2}$), the total chi-square ($\chi^2$) and the degrees of freedom (d.o.f.).
}
\end{deluxetable*}

\subsection{2.1. RXTE observations}
The PCA spectra are extracted according to the standard procedures described in \emph{RXTE} Cook Book, based on the New PCA Tools\footnote{\url{https://heasarc.gsfc.nasa.gov/docs/xte/recipes2/Overview.html}} with HEAsoft\footnote{\url{https://heasarc.gsfc.nasa.gov/docs/software/heasoft/download.html}} v6.28. The whole procedure uses the PCA calibration files\footnote{\url{https://heasarc.gsfc.nasa.gov/docs/heasarc/caldb/caldb_supported_missions.html}} v20200515. We select the data from the top layer of the best-calibrated detector PCU2 in Standard 2 mode. A dead time correction is taken into account. The latest bright-source background model is used to generate background spectra. After generating standard data products, we also apply calibration correction \texttt{pcacorr} \footnote{\url{http://www.srl.caltech.edu/personnel/javier/crabcorr/index.html}} which can reduce the systematic error that accounts for the uncertainties in the instrumental responses (\citealt{Javier2014}). We adopt a $0.1\%$ systematic error. In this work, we choose the energy band of 3.0-45.0 keV. Following our stringent screening requirements on $l$ and $f_{\mathrm{sc}} \lesssim 25 \%$, we select 9 spectra (SP1-SP9). To account for the significant flux-normalization differences between X-ray missions, we follow the previous work (\citealt{Steiner2010}) in attempt to standardize the calibration using the Crab as a reference. We opt for observations of Crab that are close to the target dates for SP1-SP9, and adopt $\Gamma=2.1$, $N=9.7 \text{ photons } s^{-1} \text{keV}^{-1}$ (\citealt{Toor1974}) as the standard values. For spectra SP1-SP8, we determine that the normalization-correction coefficient $C_{\mathrm{TS}}$ is $1.128$ and the slope difference $\Delta \Gamma_{\mathrm{TS}}$ is $0.021$. For SP9, $C_{\mathrm{TS}}$ and $\Delta \Gamma_{\mathrm{TS}}$ are 1.123, 0.018, respectively. This standardization is implemented in \texttt{XSPEC} via \texttt{crabcor}. We list the detailed information about SP1-SP9 in Table \ref{table:1}. It is worth noting that MAXI J1659 is a bright source, whose peak flux reaches about 300 mCrab (\citealt{Kalamkar2011}). \cite {Jahoda2006} put forward that the threshold of pile-up significance of \emph{RXTE/}PCA is 10,000 $\mathrm{cts} ~\mathrm{s}^{-1} ~\mathrm{PCU}^{-1}$. In other words, below this threshold, the photon pile-up is not significant. SP1-SP9 are all far below this threshold.

\begin{figure}[ht!]
\epsscale{1.0}
\centering
 \plotone{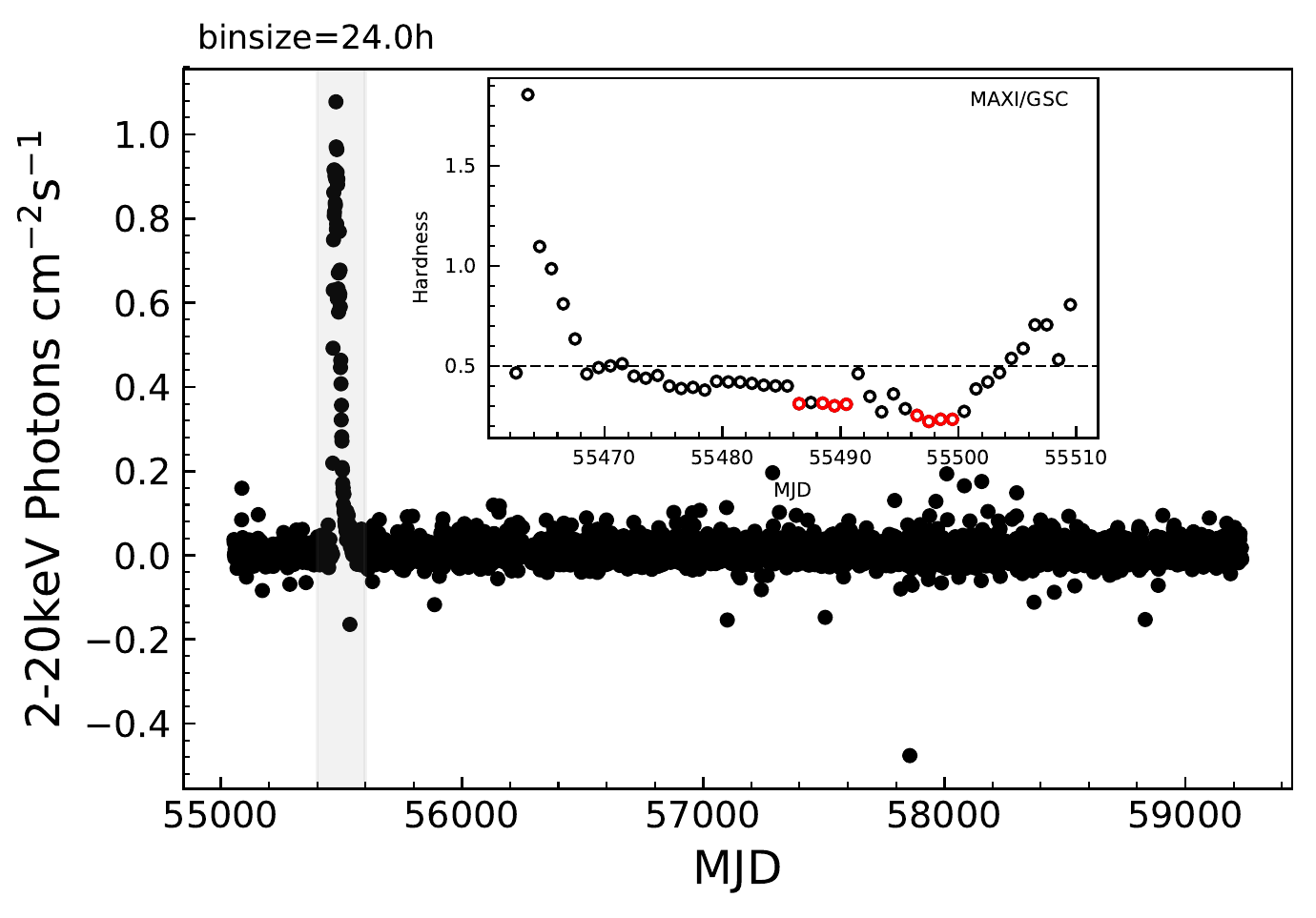}
\caption{The light curve of MAXI J1659-152 over the past 11 years as observed by \emph{MAXI}/GSC. The inset shows the evolution of the hardness ratio during the 2010 outburst. The hardness is defined as the ratio of the counts detected at 4.0-10.0 keV (hard X-ray bands) to the counts detected at 2.0-4.0 keV (soft X-ray bands) of \emph{MAXI}/GSC. The red circles in inset correspond to dates with observations} by \emph{RXTE}/PCA, with two points (SP2 and SP3) overlapping.\label{fig:1}
\end{figure}

\subsection{2.2. XMM-Newton observations} 
With \emph{RXTE}/PCA coverage beginning at 3~keV, the hydrogen column density ($N_{\mathrm{H}}$) can not be well constrained. So we turn to \emph{XMM-Newton} data which covers the low-energy band in which $N_{\mathrm{H}}$ is most prominent. We employ EPIC-pn timing-mode data during observation (ObsID 0656780601) which spanned $\sim$23 ks during the X-ray outburst. Data reduction is carried out using the Science Analysis System (SAS)\footnote{\url{https://www.cosmos.esa.int/web/xmm-newton/download-and-install-sas}} v18.0 with the latest calibration files. We follow the standard procedures and remove the influence caused by pile-up. In addition, we include $0.5\%$ systematic error as suggested by the \emph{XMM-Newton} team\footnote{\url{https://heasarc.gsfc.nasa.gov/docs/xmm/sl/epic/image/sas_cl.html}}. Crab calibration is also made for \emph{XMM-Newton} data. Implementing \texttt{TBabs*(diskbb+powerlaw)}, eventually, we get a $N_{\mathrm{H}}=3.22 \times10^{21}\mathrm{cm}^{-2}$ via fitting the spectrum over the energy range of 0.7-12.0 keV with a reduced chi-square $\chi_{v}^{2}=1.179~(2451.8 / 2080.0)$. In subsequent work, all $N_{\mathrm{H}}$ are fixed to this value.

\section{3. SPECTRAL ANALYSIS AND RESULTS}
\label{section:3}
\subsection{3.1. The Non-Relativistic Model}
\begin{figure}[ht!]
\epsscale{1.0}
\centering
 \plotone{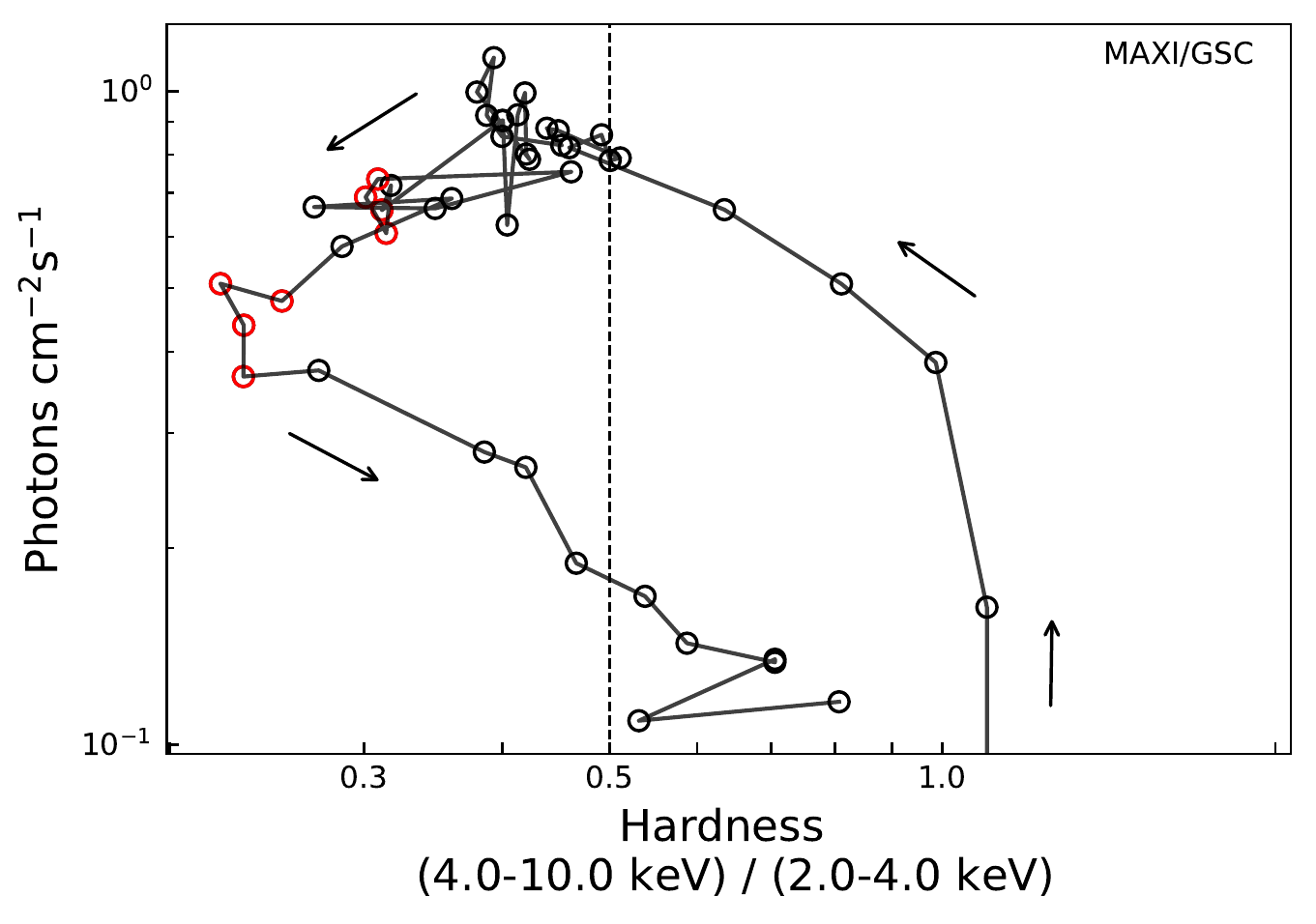}
\caption{The hardness-intensity diagram of MAXI J1659-152 by \emph{MAXI}/GSC observations. The hardness is defined as the ratio of the counts detected at 4.0-10.0 keV to the counts detected at 2.0-4.0 keV of \emph{MAXI}/GSC. The red circles above represent the observations by \emph{RXTE}/PCA that we analyze in this paper, again with two points (SP2 and SP3) overlapping.\label{fig:2}}
\end{figure}

As shown in Figure \ref{fig:1}, \emph{MAXI}/GSC has been monitoring MAXI J1659 for up to eleven years. The horizontal axis represents MJD and the vertical axis represents the flux in 2.0-20.0 keV detected by \emph{MAXI}/GSC. As can be seen from Figure \ref{fig:1}, MAXI J1659 has exhibited just one major outburst and shown no other signs of activity. As displayed in Figure \ref{fig:2}, the hardness-intensity diagram (HID) shows a classical ``q''-like shape, which is a typical for an outbursting black hole LMXB.

\setlength{\tabcolsep}{3mm}
\begin{deluxetable*}{cccrrrrlcc}
\centering
\tablecaption{The best-fitting parameters for SP1-SP9 with crabcor*TBabs*(simpl*kerrbb2) ~($\alpha=0.1$)} \label{table:3}
\tabletypesize{\scriptsize}
\tablehead{
\colhead{Group} &\colhead{Spec.} & \colhead{ObsID} & \multicolumn{2}{c}{SIMPL} & \multicolumn{2}{c}{KERRBB2} & \colhead{$\chi^2_\nu$} & \colhead{$\chi^2$(d.o.f.)} & \colhead{$L/ L_{\mathrm{Edd}}$}\\
\cline{4-7}
\colhead{} &\colhead{} & \colhead{} &\colhead{$\Gamma$} & \colhead{$f_{\rm sc}$} & \colhead{$a_{*}$} & \colhead{$\dot{M}$} 
} 
\startdata
&SP1	&	95118-01-03-00	&	2.39 	$\pm$	0.02 	&  0.189	$\pm$	0.003 	&	0.163 	$\pm$	0.03  &	1.36 	$\pm$	0.05    &  1.141 & 77.62 (68) &	0.109\\
&SP2	&	95118-01-05-00	&	2.40	$\pm$	0.02 	&  0.227 	$\pm$	0.003 	&	0.146 	$\pm$	0.03 &	1.33 	$\pm$	0.04    &  1.422 & 96.71 (68) &	0.104\\
&SP3	&	95118-01-05-01	&	2.31 	$\pm$	0.02 	&  0.191 	$\pm$	0.003 	&	0.217 	$\pm$	0.02 &	1.18 	$\pm$	0.03 	   &  1.085 & 73.76 (68) &	0.097 \\
&SP4	&	95118-01-06-00	&	2.35 	$\pm$	0.02 	&  0.166 	$\pm$	0.003 	&	0.205 	$\pm$	0.02 &	1.24	$\pm$	0.03    &  0.961 & 65.34 (68) &	0.102\\
&SP5	&	95118-01-07-01	&	2.37 	$\pm$	0.02 	&  0.245 	$\pm$	0.003 	&	0.234	$\pm$	0.03 &	1.15 	$\pm$	0.04    &  1.171 & 79.63 (68) &	0.096\\
&SP6	&	95118-01-13-00	&	2.31 	$\pm$	0.02 	&  0.237 	$\pm$	0.003 	&	0.263 	$\pm$	0.04 &	0.84 	$\pm$	0.05    &  0.834 & 56.71 (68) &	0.074\\
&SP7	&	95118-01-14-00	&	2.25 	$\pm$	0.05 	&  0.152 	$\pm$	0.004 	&	0.277 	$\pm$	0.06 &	0.86 	$\pm$	0.07    &  0.852 & 57.94 (68) &	0.056\\
&SP8	&	95118-01-15-00	&	2.34 	$\pm$	0.03 	&  0.203 	$\pm$	0.005 	&	0.143	$\pm$	0.07 &	0.96 	$\pm$	0.08    &  1.133 & 77.02 (68) &	0.076\\
Group 1&SP9	&	95118-01-15-01	&	2.34 	$\pm$	0.03 	&  0.215 	$\pm$	0.004 	&	0.071 	$\pm$	0.09 &	1.01 	$\pm$	0.09    &  0.734 & 49.92 (68) &	0.076\\
\cline{2-10}
&SP1	&	95118-01-03-00	&	2.38 	$\pm$	0.01 	&  0.188	$\pm$	0.002 	&	 		  &	1.33 	$\pm$	0.02    &   &  & 0.096	\\
&SP2	&	95118-01-05-00	&	2.38	$\pm$	0.01 	&  0.225 	$\pm$	0.002 	&	 		 &	1.27 	$\pm$	0.02    &   &  &	0.099\\
&SP3	&	95118-01-05-01	&	2.33 	$\pm$	0.01 	&  0.194 	$\pm$	0.002 	&	 		 &	1.23 	$\pm$	0.02 	   &   &  &	0.096 \\
&SP4	&	95118-01-06-00	&	2.37 	$\pm$	0.02 	&  0.167 	$\pm$	0.002 	&	 		 &	1.27	$\pm$	0.02    &   &  &	0.099\\
&SP5	&	95118-01-07-01	&	2.39 	$\pm$	0.01 	&  0.250 	$\pm$	0.003 	&     0.184	$\pm$0.02 &	1.23 	$\pm$	0.02    &  1.049 & 650.65 (620) & 0.096\\
&SP6	&	95118-01-13-00	&	2.34 	$\pm$	0.01 	&  0.241 	$\pm$	0.003 	&	 		 &	0.93 	$\pm$	0.02    &  &  &	0.072\\
&SP7	&	95118-01-14-00	&	2.31 	$\pm$	0.03 	&  0.157 	$\pm$	0.004 	&	 		 &	0.97 	$\pm$	0.02    &  &  &	0.075\\
&SP8	&	95118-01-15-00	&	2.33 	$\pm$	0.02 	&  0.202 	$\pm$	0.004 	&			 &	0.92 	$\pm$	0.02    &  &  &	0.071\\
&SP9	&	95118-01-15-01	&	2.31 	$\pm$	0.02 	&  0.212 	$\pm$	0.004 	&	 		 &	0.89 	$\pm$	0.02    &  & &	0.069\\
\hline
\hline
&SP1	&	95118-01-03-00	&	2.39 	$\pm$	0.02 	&  0.191 	$\pm$	0.003 	&	-0.872 	$\pm$	0.05 &	4.55 	$\pm$	0.14   &  1.175& 79.92 (68) &	0.263\\
&SP2	&	95118-01-05-00	&	2.39 	$\pm$	0.02 	&  0.229 	$\pm$	0.003 	&	-0.888 	$\pm$	0.05 &	4.40 	$\pm$	0.14   &  1.458 & 99.16 (68) &	0.252 \\
&SP3	&	95118-01-05-01	&	2.31 	$\pm$	0.02 	&  0.194 	$\pm$	0.003 	&	-0.778 	$\pm$	0.05 &	3.96 	$\pm$	0.13   &  1.113 & 75.70 (68) &	0.236 \\
&SP4	&	95118-01-06-00	&	2.36 	$\pm$	0.02 	&  0.169 	$\pm$	0.002 	&	-0.840 	$\pm$	0.05 &	4.25	$\pm$	0.13   &  0.969 & 65.88 (68) &	0.248\\
&SP5	&	95118-01-07-01	&	2.37 	$\pm$	0.02 	&  0.249 	$\pm$	0.003 	&	-0.737	$\pm$	0.06 &	3.86 	$\pm$	0.14   &  1.193 & 81.13 (68) &	0.234 \\
&SP6	&	95118-01-13-00	&	2.30 	$\pm$	0.02 	&  0.239 	$\pm$	0.004 	&	-0.635 	$\pm$	0.08 &	2.73 	$\pm$	0.16   &  0.839 & 57.07 (68) &	0.172\\
&SP7	&	95118-01-14-00	&	2.24 	$\pm$	0.05 	&  0.154 	$\pm$	0.004 	&	-0.620 	$\pm$	0.11 &       2.81	$\pm$	0.22   &  0.854 & 58.04 (68) &	0.178 \\
&SP8	&	95118-01-15-00	&	2.34 	$\pm$	0.03 	&  0.205 	$\pm$	0.005 	&	-0.874 	$\pm$	0.10 &	3.15 	$\pm$	0.22   &  1.138 & 77.36 (68) &	0.182\\
Group 2&SP9	&	95118-01-15-01	&	2.34 	$\pm$	0.03 	&  0.217 	$\pm$	0.004 	&	-0.959 	$\pm$	0.10 &	3.24 	$\pm$	0.22   &  0.736 & 50.03 (68) &	0.182\\
\cline{2-10}
&SP1	&	95118-01-03-00	&	2.37 	$\pm$	0.02 	&  0.189 	$\pm$	0.002 	&	 		 &	4.42 	$\pm$	0.07    &   &  &	 0.250\\
&SP2	&	95118-01-05-00	&	2.38 	$\pm$	0.01 	&  0.227 	$\pm$	0.002 	&	 		 &	4.23	$\pm$	0.06    &   &  &	0.239\\
&SP3	&	95118-01-05-01	&	2.33 	$\pm$	0.02 	&  0.196 	$\pm$	0.002 	&	 		 &	4.08 	$\pm$	0.06    &   &  &	0.231\\
&SP4	&	95118-01-06-00	&	2.35 	$\pm$	0.02 	&  0.168 	$\pm$	0.002 	&	 		 &	4.21	$\pm$	0.06    &   &  &	0.238\\
&SP5	&	95118-01-07-01	&	2.39 	$\pm$	0.01 	&  0.252 	$\pm$	0.003 	&	-0.826$\pm$	0.02 &	4.09 	$\pm$	0.06 	  &  1.066 & 661.14 (620) & 0.231\\
&SP6	&	95118-01-13-00	&	2.34 	$\pm$	0.01 	&  0.244 	$\pm$	0.003 	&	 		 &	3.12 	$\pm$	0.05   &   &  &	0.176\\
&SP7	&	95118-01-14-00	&	2.32 	$\pm$	0.03 	&  0.160 	$\pm$	0.004 	&	 		 &	3.25 	$\pm$	0.05 	  &   &  &	0.184\\
&SP8	&	95118-01-15-00	&	2.32 	$\pm$	0.02 	&  0.204 	$\pm$	0.004 	&	 		 &     3.05 	$\pm$	0.05 	  &  &  &	0.172\\
&SP9	&	95118-01-15-01	&	2.31 	$\pm$	0.02 	&  0.214 	$\pm$	0.004 	&			 &     2.96 	$\pm$	0.05 	  &   & &	0.167\\
\enddata
{Notes.\\
In columns 3-10, we show undermentioned information: the observation ID (ObsID), the dimensionless photon index of power-law ($\Gamma$), the scattered fraction ($f_{\rm sc}$), the dimensionless spin parameter ($a_{*}$), the effective mass accretion rate of the disk in units of 10$^{18}$ g s $^{-1}$ ($\dot{M}$), the reduced chi-square ($\chi_{\nu}^{2}$), the total chi-square ($\chi^2$) and the degrees of freedom (d.o.f.), the bolometric Eddington-scaled luminosities ($L(a_{*}, \dot{M})/ L_{\mathrm{Edd}}$) ~(where $L_{\mathrm{Edd}}= 1.3 \times 10^{38}\left(M / M_{\odot}\right)$ erg $\mathrm{s}^{-1}$, see \citealt{Shapiro1983}).\\Single horizontal lines are used to distinguish between simultaneous fitting and independent fitting, above the single horizontal line are the results of independent fitting of 9 spectra, and below the single horizontal line are the results of simultaneous fitting of 9 spectra. }The two sets of system parameters: $M=5.7 M_{\odot}$, $i=70.0^{\circ}$, $D=6 \pm 2$~kpc (group 1) and $M=4.9M_{\odot}$, $i=80.0^{\circ}$, $D=6 \pm 2$~kpc (group 2) are separated with double horizontal lines. Above the double horizontal lines is group 1, below lines is group 2.
\end{deluxetable*}

\setlength{\tabcolsep}{3mm}
\begin{deluxetable*}{cccrrrrlcc}
\centering
\tablecaption{The best-fitting parameters for SP1-SP9 with \texttt{crabcor*TBabs*(simpl*kerrbb2)} ~($\alpha=0.01$)} \label{table:4}
\tabletypesize{\scriptsize}
\tablehead{
\colhead{Group}&\colhead{Spec.} & \colhead{ObsID} & \multicolumn{2}{c}{simpl} & \multicolumn{2}{c}{kerrbb2}  & \colhead{$\chi^2_\nu$} & \colhead{$\chi^2$(d.o.f.)} & \colhead{$L/ L_{\mathrm{Edd}}$}\\
\cline{4-7}
\colhead{} &\colhead{} & \colhead{} &\colhead{$\Gamma$} & \colhead{$f_{\rm sc}$} & \colhead{$a_{*}$} & \colhead{$\dot{M}$} 
} 
\startdata
&SP1	&	95118-01-03-00	&	2.38 	$\pm$	0.02 	&  0.188 	$\pm$	0.003 	&	0.210 	$\pm$	0.02 &	1.31 	$\pm$	0.03    &  1.148 & 78.09 (68) &	0.107\\
&SP2	&	95118-01-05-00	&	2.40 	$\pm$	0.02 	&  0.228 	$\pm$	0.003 	&	0.177 	$\pm$	0.03 &	1.30 	$\pm$	0.05    &  1.430 & 97.25 (68) &	0.104\\
&SP3	&	95118-01-05-01	&	2.31 	$\pm$	0.02 	&  0.192 	$\pm$	0.003 	&	0.237 	$\pm$	0.02 &	1.17 	$\pm$	0.04    &  1.080 & 73.45 (68) &	0.097\\
&SP4	&	95118-01-06-00	&	2.35 	$\pm$	0.02 	&  0.166 	$\pm$	0.003 	&	0.218 	$\pm$	0.02 &	1.23	$\pm$	0.03    &  0.956 & 65.03 (68) &	0.102\\
& SP5	&	95118-01-07-01	&	2.37 	$\pm$	0.02 	&  0.246 	$\pm$	0.003 	&	0.254	$\pm$	0.03 &	1.14 	$\pm$	0.04    &  1.172 & 79.70 (68) &	0.096\\
&SP6	&	95118-01-13-00	&	2.30 	$\pm$	0.02 	&  0.236 	$\pm$	0.004 	&	0.310 	$\pm$	0.03 &	0.80 	$\pm$	0.03    &  0.838 & 56.98 (68) &	0.070\\
&SP7	&	95118-01-14-00	&	2.24 	$\pm$	0.05 	&  0.152 	$\pm$	0.005 	&	0.314 	$\pm$	0.03 &	0.83 	$\pm$	0.05    &  0.852 & 57.95 (68) &	0.073\\
&SP8	&	95118-01-15-00	&	2.34 	$\pm$	0.03 	&  0.203 	$\pm$	0.004 	&	0.165 	$\pm$	0.07 &	0.95 	$\pm$	0.08    &  1.137 & 77.30 (68) &	0.076\\
Group 1&SP9	&	95118-01-15-01	&	2.34 	$\pm$	0.03 	&  0.214 	$\pm$	0.004 	&	0.132 	$\pm$	0.05 &	0.96 	$\pm$	0.06    & 0.733 & 49.84 (68) &	0.075\\
\cline{2-10}
&SP1	&	95118-01-03-00	&	2.37 	$\pm$	0.02 	&  0.186 	$\pm$	0.002 	&	 		 	        &               1.28    $\pm$	0.01                                     &   &  &	0.102\\
&SP2	&	95118-01-05-00	&	2.37 	$\pm$	0.01 	&  0.223 	$\pm$	0.002 	&	 		 	        &               1.23    $\pm$	0.01                                     &   &  &	0.098\\
&SP3	&	95118-01-05-01	&	2.32 	$\pm$	0.02 	&  0.193	$\pm$	0.002	&	 		 	        &               1.18    $\pm$	0.01                                     &  &  &	0.094\\
&SP4	&	95118-01-06-00	&	2.35 	$\pm$	0.02 	&  0.165 	$\pm$	0.002 	&	 		                &            	1.22    $\pm$	0.01                                     &   &  &	0.097\\
&SP5	&	95118-01-07-01	&	2.39 	$\pm$	0.01 	&  0.249 	$\pm$	0.003 	&	0.224$\pm$	0.01 &	         1.18    $\pm$	0.01    &  1.048 & 649.55 (620) &	0.094\\
&SP6	&	95118-01-13-00	&	2.34 	$\pm$	0.01 	&  0.241 	$\pm$	0.003 	&	 		                &                0.90    $\pm$	0.01                                     &   &  &	0.072\\
&SP7	&	95118-01-14-00	&	2.30 	$\pm$	0.03 	&  0.157 	$\pm$	0.004 	&	 		 	        &                0.94    $\pm$	0.01                                     &   &  &	0.075\\
&SP8	&	95118-01-15-00	&	2.31 	$\pm$	0.02 	&  0.200 	$\pm$	0.004 	&	 		 	        &                0.88    $\pm$	0.01                                      &   &  &	0.070\\
&SP9	&	95118-01-15-01	&	2.30 	$\pm$	0.02 	&  0.211 	$\pm$	0.004 	&	 		 	        &                0.85    $\pm$	0.01                                      &   &  &	0.068\\
\hline
\hline
&SP1	&	95118-01-03-00	&	2.39 	$\pm$	0.02 	&  0.191 	$\pm$	0.003 	&	-0.684 	$\pm$	0.04 &	4.19 	$\pm$	0.12    &  1.170 & 79.53 (68) &	0.259 \\
&SP2	&	95118-01-05-00	&	2.40 	$\pm$	0.02 	&  0.230 	$\pm$	0.003 	&	-0.731 	$\pm$	0.06 &	4.13 	$\pm$	0.15    &  1.451 & 98.66 (68) &	0.250\\
&SP3	&	95118-01-05-01	&	2.32 	$\pm$	0.02 	&  0.194 	$\pm$	0.003 	&	-0.641 	$\pm$	0.05 &	3.74 	$\pm$	0.14    &  1.107 & 75.27 (68) &	0.234\\
&SP4	&	95118-01-06-00	&	2.36 	$\pm$	0.02 	&  0.168 	$\pm$	0.003 	&	-0.657 	$\pm$	0.04 &	3.92	$\pm$	0.11    &  0.964 & 65.58 (68) &	0.244\\
&SP5	&	95118-01-07-01	&	2.37 	$\pm$	0.02 	&  0.248 	$\pm$	0.004 	&	-0.584	$\pm$	0.04 &	3.59 	$\pm$	0.12 	  &  1.191 & 80.97 (68) &	0.230\\
&SP6	&	95118-01-13-00	&	2.30 	$\pm$	0.02 	&  0.239 	$\pm$	0.004 	&	-0.530 	$\pm$	0.07 &	2.60 	$\pm$	0.14   &  0.838 & 57.00 (68) &	0.170\\
&SP7	&	95118-01-14-00	&	2.24 	$\pm$	0.05 	&  0.154 	$\pm$	0.004 	&	-0.519 	$\pm$	0.09 &	2.68 	$\pm$	0.19 	  &  0.853 & 58.02 (68) &	0.176\\
&SP8	&	95118-01-15-00	&	2.34 	$\pm$	0.03 	&  0.205 	$\pm$	0.005 	&	-0.761 	$\pm$	0.10 &      3.01 	$\pm$	0.21 	  &  1.136 & 77.27 (68) &	0.181\\
Group 2&SP9	&	95118-01-15-01	&	2.34 	$\pm$	0.03 	&  0.217 	$\pm$	0.004 	&	-0.835 	$\pm$	0.12 &      3.09 	$\pm$	0.24 	  &  0.735 & 49.99 (68) &	0.181\\
\cline{2-10}
&SP1	&	95118-01-03-00	&	2.38 	$\pm$	0.02 	&  0.189 	$\pm$	0.002 	&	 		 &	4.12 	$\pm$	0.05    &   &  &	0.247 \\
&SP2	&	95118-01-05-00	&	2.38 	$\pm$	0.01 	&  0.226 	$\pm$	0.002 	&	 		 &	3.93	$\pm$	0.05    &   &  &	0.236\\
&SP3	&	95118-01-05-01	&	2.32 	$\pm$	0.02 	&  0.195 	$\pm$	0.002 	&	 		 &	3.78	$\pm$	0.05    &   &  &	0.227\\
&SP4	&	95118-01-06-00	&	2.36 	$\pm$	0.02 	&  0.168 	$\pm$	0.002 	&	 		 &	3.92	$\pm$	0.05    &   &  &	0.235\\
&SP5	&	95118-01-07-01	&	2.39 	$\pm$	0.01 	&  0.251 	$\pm$	0.003 	&	-0.659$\pm$	0.02 &	3.78 	$\pm$	0.05 	  &  1.057 & 655.57 (620) & 0.227\\
&SP6	&	95118-01-13-00	&	2.34 	$\pm$	0.01 	&  0.243 	$\pm$	0.003 	&	 		 &	2.87 	$\pm$	0.04   &   &  &	0.172\\
&SP7	&	95118-01-14-00	&	2.30 	$\pm$	0.03 	&  0.159 	$\pm$	0.004 	&	 		 &	2.99 	$\pm$	0.04 	  &   &  &	0.179\\
&SP8	&	95118-01-15-00	&	2.31 	$\pm$	0.02 	&  0.202 	$\pm$	0.004 	&	 		 &     2.80 	$\pm$	0.04 	  &  &  &	0.168\\
&SP9	&	95118-01-15-01	&	2.30 	$\pm$	0.02 	&  0.213 	$\pm$	0.003 	&			 &     2.72 	$\pm$	0.04 	  &   & &	0.163\\
\enddata
{Notes.\\
In columns 3-10, we show the following information: the observation ID (ObsID), the dimensionless photon index of power-law ($\Gamma$), the scattered fraction ($f_{\rm sc}$), the dimensionless spin parameter ($a_{*}$), the mass accretion rate through the disk in units of 10$^{18}$ g s $^{-1}$ ($\dot{M}$), the reduced chi-square ($\chi_{\nu}^{2}$), the total chi-square ($\chi^2$) and the degrees of freedom (d.o.f.), the luminosity ($l$). \\Single horizontal lines are used to distinguish between simultaneous fitting and independent fitting, above the single horizontal line are results of independent fitting of 9 spectra, and below the single horizontal line are the results of simultaneous fitting of 9 spectra. }The two sets of system parameters: $M=5.7 M_{\odot}$, $i=70.0^{\circ}$, $D=6 \pm 2$~kpc (group 1) and $M=4.9M_{\odot}$, $i=80.0^{\circ}$, $D=6 \pm 2$~kpc (group 2) are separated with double horizontal lines. Above the double horizontal double lines is group 1, below the double lines is group 2.
\end{deluxetable*}
 
\begin{figure}
\epsscale{1.0}
\centering
 \plotone{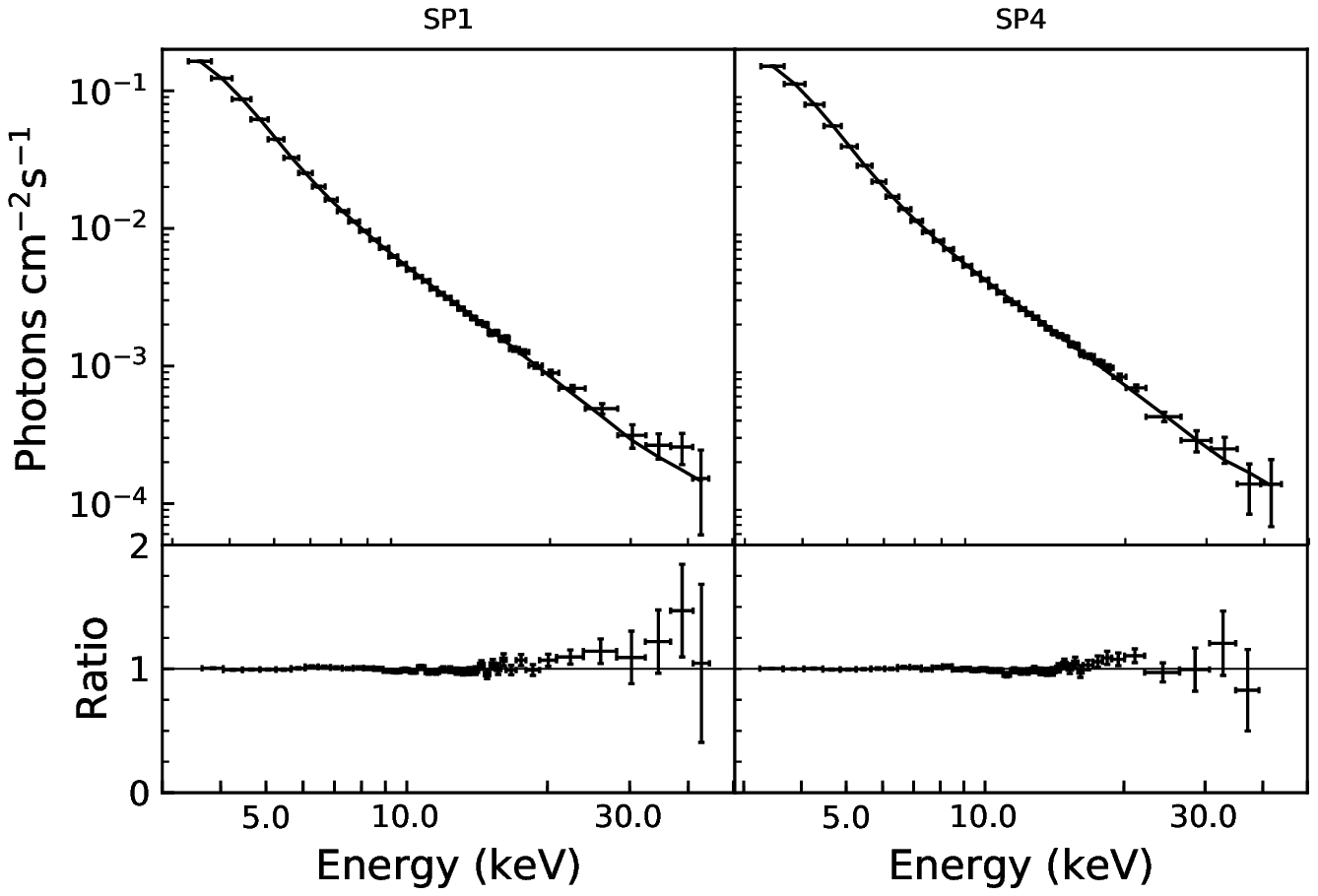}
\caption{Illustrative spectral fits (non-relativistic) for SP1 (ObsID 95118-01-03-00) and SP4 (ObsID 95118-01-06-00). Data have been rebinned for visual clarity. \label{fig:3}}
\end{figure}

Firstly, we present the non-relativistic case, which consists of multiple blackbody disk components \texttt{diskbb} (\citealt{Mitsuda1984}; \citealt{Makishima1986}). The composite  non-relativistic model \texttt{crabcor*TBabs*(simpl*diskbb)} is used to fit the data for the 9 screened \emph{RXTE}/PCA spectral. The model \texttt{crabcor} and model \texttt{simpl} have been mentioned earlier (see Section \ref{section:2} and Section \ref{section:1} respectively). The photoionization cross-sections of the interstellar medium (ISM) in model \texttt{TBabs} are based on \cite {Verner1996} and the abundances are based on \cite{Wilms2000}, and we set $N_{\mathrm{H}}$ to $3.22 \times 10^{21} \mathrm{cm}^{-2}$ (see Section \ref{section:2}). The best-fitting results of SP1-SP9 are listed in Table \ref{table:2}. As can be seen, these 9 spectra are well fitted and the reduced chi-square $\chi_{\nu}^2$ of SP1-SP9 are basically maintained in the vicinity of 1. Figure \ref{fig:3} shows the fitting of SP1 and SP4 as representatives. We can discover that the model fitted well without any significant residual.

\begin{figure}[ht!]
\epsscale{1.0}
\centering
\plotone{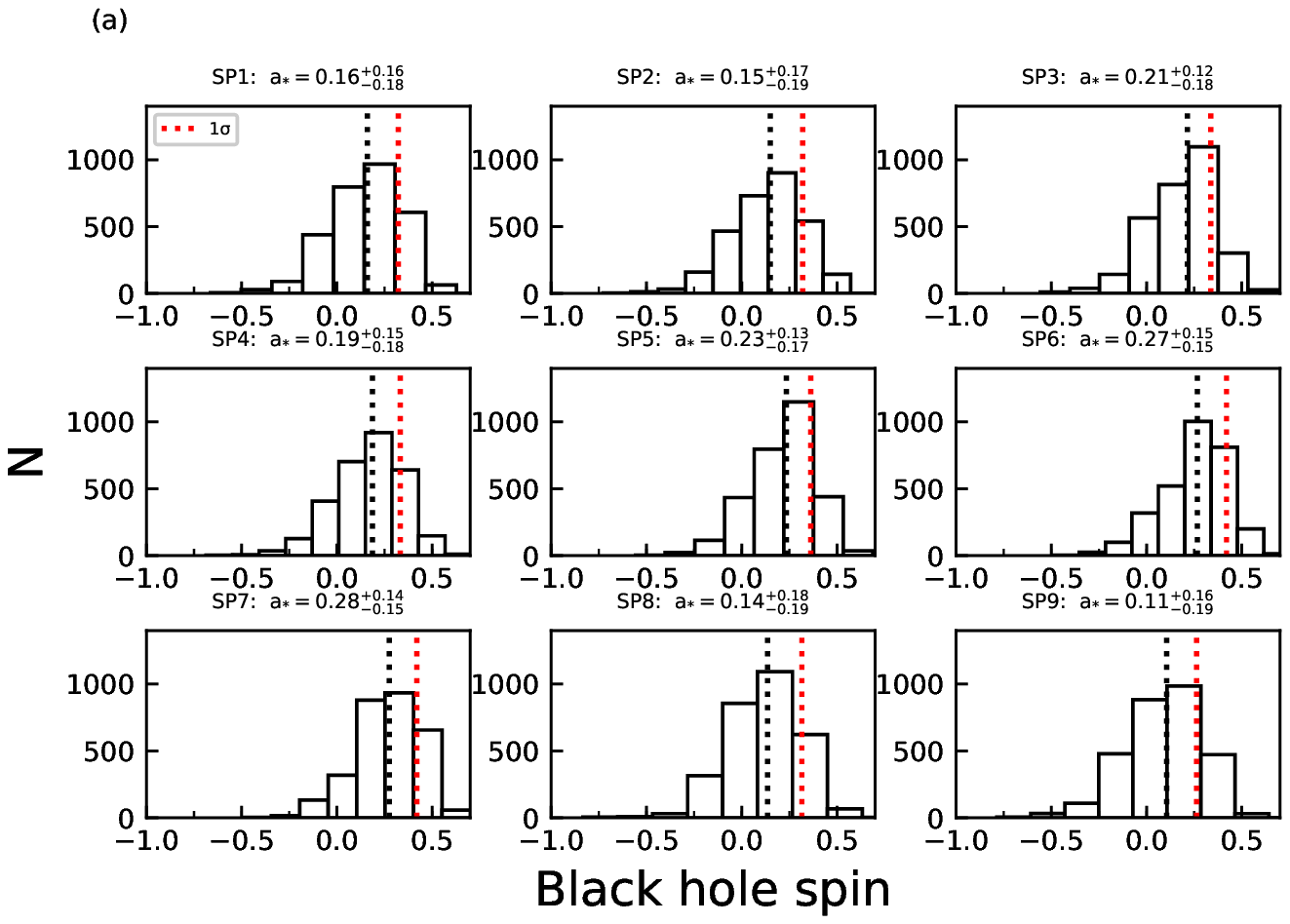}
\plotone{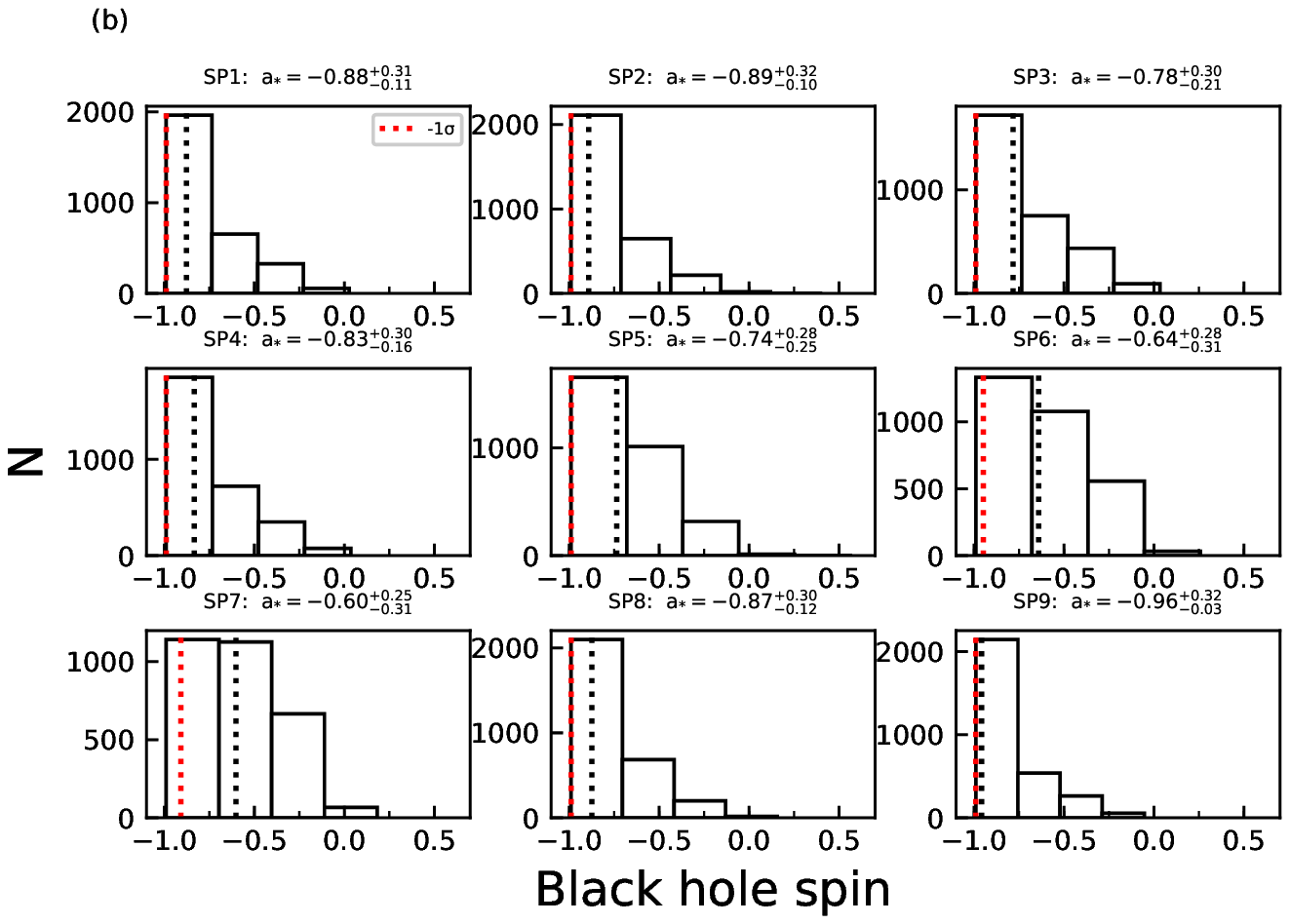}
\caption{(a). The results of MC error analysis of $a_{*}$ for SP1-SP9 for group 1. (b). The analogous results for group 2.  For both (a) and (b), black dotted lines represent the center value of $a_{*}$, and red dotted lines represent the $\pm 68.3\% $ ($\pm$ 1$\sigma$) lower and upper-limits between groups 1 and 2, respectively. \label{fig:4}}
\end{figure}

\subsection{3.2. The Relativistic Model}
We next move to the fully-relativistic accretion-disk model 
\texttt{kerrbb2} (\citealt{McClintock2006}). \texttt{kerrbb2} merges two different disk models, \texttt{bhspec} and \texttt{kerrbb}. Specifically, \texttt{bhspec} is used to determine the value of the spectral hardening factor $f \equiv T_{\mathrm{col}} / T_{\mathrm{eff}}$ (also known as the color correction factor, \citealt{Davis2005}), while \texttt{kerrbb} employs ray-tracing computations to model the disk (\citealt{Li2005}). We first use \texttt{bhspec}\footnote{\url{http://people.virginia.edu/~swd8g/xspec.html}} to compute spectral-hardening look-up tables according to two representative values of the viscosity parameter ($\alpha=0.1$ and $\alpha=0.01$). It is worth noting that the calculations with the model \texttt{BHSPEC} are made for spins ranging from -1 to 1. During kerrbb2 fitting, the $f$-table is read in and used to automatically set $f$'s value;  otherwise the performance is identical to the original \texttt{kerrbb}. The relativistic model can be expressed as \texttt{crabcor*TBabs*(simpl*kerrbb2)}.  
By default we adopt $\alpha=0.1$ and note that $\alpha=0.01$ returns {\em slightly} larger values for spin.  

We adopt the revised system parameters from improved by \cite{Torres2021}, and consider inclination angles of $70^{\circ}$ and $80^{\circ}$. The mass is restricted to $5.7 \pm 1.8~\mathrm{M}_{\odot}$ (1$\sigma$) and $4.9 \pm 1.6~\mathrm{M}_{\odot}$ (1$\sigma$) respectively. Following \cite {Torres2021}, when the distance is $D=6 \pm 2$\text{ ~kpc}, the center value of the measured mass falls within the range of 4.3-6.0$~\mathrm{M}_{\odot}$ calculated based on formula 5 by \cite {Yamaoka2012}. In addition, this distance is also consistent with \cite {Kuulkers2013}. Therefore, we adopt the distance of $D=6 \pm 2$$ \text{ ~kpc}$ in this paper. 

The self-irradiation of the disk (rflag=1) and the effect of limb-darkening (lflag=1) are taken into account. And we set the torque at the inner boundary of the disk to be zero (eta=0). For both bracketing sets of system parameters: $M=5.7 \pm 1.8~\mathrm{M}_{\odot}$ (1$\sigma$), $i=70.0^{\circ}$, $D=6 \pm 2$~kpc (hereafter group 1) and $M=4.9 \pm 1.6 ~\mathrm{M}_{\odot}$ (1$\sigma$), $i=80.0^{\circ}$, $D=6 \pm 2$~kpc (hereafter group 2), we show our results in Table \ref{table:3}. Above the double horizontal lines is group 1, below the double horizontal lines is group 2. Group 1 (70$^\circ$) would have MAXI J1659 with a moderate positive spin, whereas group 2 (80$^\circ$) finds an extreme retrograde spin as most likely. 

Although in this paper, we focus upon results for $\alpha=0.1$ as our default, for the alternative case of $\alpha=0.01$ we have performed the same calculations. Those results are in Table \ref{table:4}. The errors that appear in Table \ref{table:3} and \ref{table:4} are only due to the statistical uncertainties estimated by \texttt{XSPEC} for $90\%$ confidence. Next, we will discuss the error from  uncertainties of input parameters: $M$, $i$, $D$ in detail.

\begin{figure}[ht!]
\epsscale{1.0}
\centering
\plotone{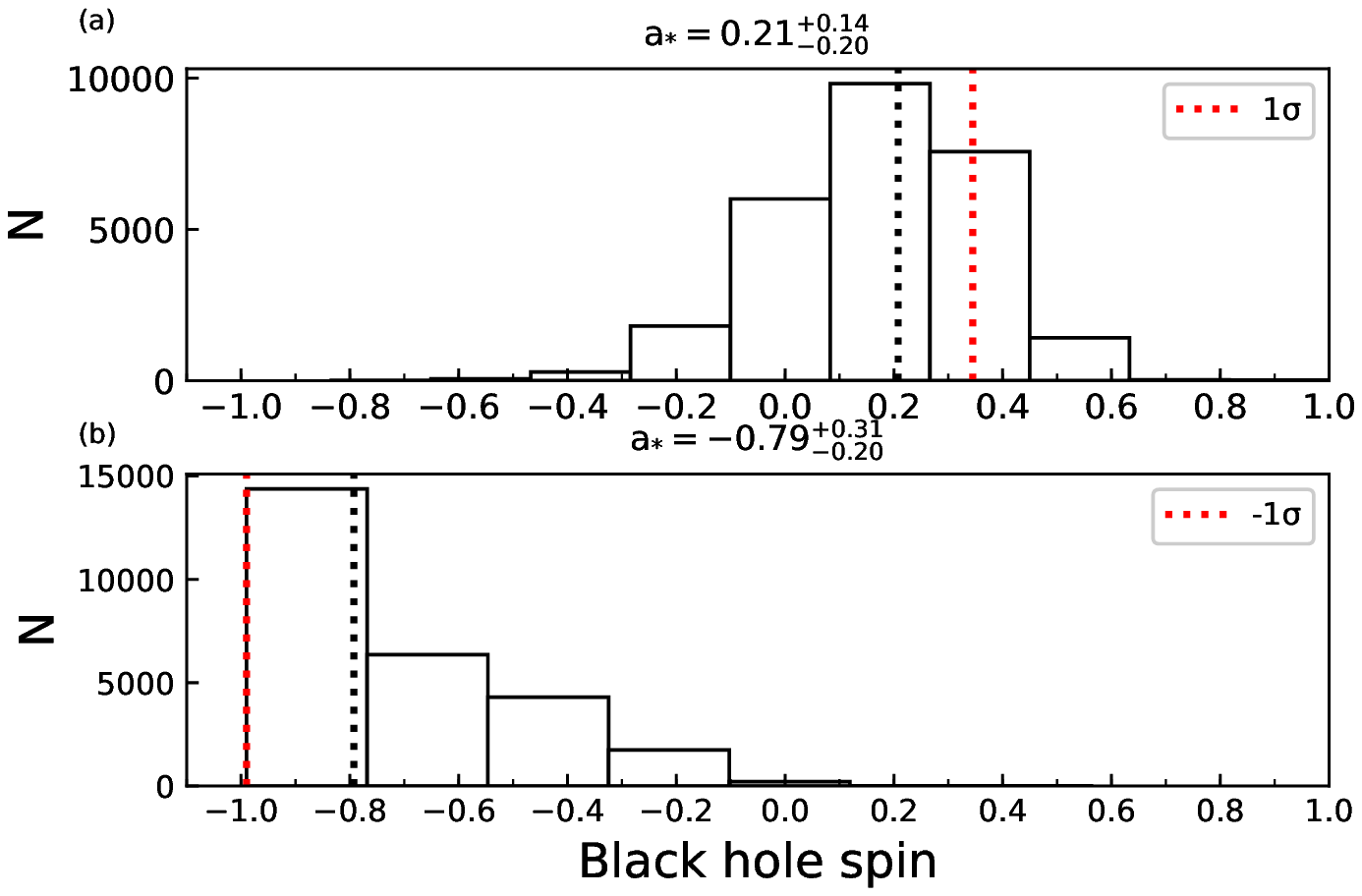}
\caption{(a). The total histogram of $a_{*}$ for SP1-SP9 for group 1, which consists of 27000 data points. (b). The analogous histogram for group 2. For both (a) and (b), black dotted lines represent the center value of $a_{*}$, and red dotted lines represent the $\pm 68.3\% $ ($\pm$ 1$\sigma$) lower and upper-limits between groups 1 and 2, respectively. \label{fig:5}}
\end{figure}

\subsection{3.3. Error Analysis}
The dominant source of uncertainty in CF is the uncertainty of the input parameters: $M$, $i$, $D$. In order to determine those uncertainties, Monte Carlo (MC) methods are often used (see, e.g., \citealt{Liu2008}; \citealt{Gou2009}). We follow these works, but differ in that, unlike those, here there are two distinct sets of input system parameters (group 1 and group 2). We first consider these groups separately.  

For each individual spectrum in SP1-SP9, assuming that randomly generated data points are independent of each other and obey the Gaussian distribution, we have generated 3000 data points of M and D respectively. $i=70.0^{\circ}$ ($i=80.0^{\circ}$), and these 3000 data points constituted 3000 data sets as input parameters. We next compute the look-up tables of $f$ for these 3000 data sets. Lastly, we use the composite model \texttt{crabcor*TBabs*(simpl*kerrbb2)} (see Section \ref{section:2}) to fit 3000 data sets to obtain the histogram of spin, so as to determine the errors. The MC-determined error analysis of each spectrum is shown in Figure \ref{fig:4}. The histograms of $a_{*}$ for SP1-SP9 is shown in Figure \ref{fig:5}. As shown in Figure \ref{fig:5}, group 1 with higher-mass and lower inclination yields spin $a_{*}=0.21_{-0.20}^{+0.14}$ (1$\sigma$); group 2 gives $a_{*}=-0.79_{-0.20}^{+0.31}$ (1$\sigma$).

 We now combine groups 1 and 2 in order to establish a net constraint on spin. We are interested in assessing the lower and upper bounds on spin which are mostly constrained by groups 2 and 1, respectively. We mark the appropriate 1-$\sigma$ limit for each spectrum in Figure \ref{fig:4}, and for the composite result in Figure \ref{fig:5}. In total, we find that the spin of MAXI J1659 is poorly constrained, with an allowable range $-1 < a_{*} \lesssim 0.35$ in 1 sigma interval. The 90\% upper limit on spin is 0.44 from group 1.

\setlength{\tabcolsep}{3mm}
\begin{deluxetable*}{crrrrccc}
\centering
\tablecaption{The best-fitting parameters for SP1 with \texttt{crabcor*TBabs*(simpl*kerrbb2)} ~($\alpha=0.1$)} \label{table:5}
\tablewidth{700pt}
\tabletypesize{\scriptsize}
\tablehead{
\colhead{$N_{\mathrm{H}}$} & \multicolumn{2}{c}{simpl} & \multicolumn{2}{c}{kerrbb2} & \colhead{$\chi^2_\nu$} & \colhead{$\chi^2$(d.o.f.)} & \colhead{$L/ L_{\mathrm{Edd}}$}\\
\cline{2-5}
\colhead{($\times10^{22} \mathrm{cm}^{-2}$)}  &\colhead{$\Gamma$} & \colhead{$f_{\rm sc}$} & \colhead{$a_{*}$} & \colhead{$\dot{M}$} 
} 
\startdata
0.200	&	2.38 	$\pm$	0.02 	&  0.189 	$\pm$	0.003 	&	0.210 	$\pm$	0.02 &	1.29	$\pm$	0.03    &  1.118     & 76.03 (68) &	0.106\\
0.322	&	2.39 	$\pm$	0.02 	&  0.189 	$\pm$	0.003 	&	0.163	$\pm$	0.03 &	1.36	$\pm$	0.05    &  1.141     & 77.62 (68) &	0.109\\
0.400	&	2.39 	$\pm$	0.02 	&  0.189 	$\pm$	0.003 	&	0.152	$\pm$	0.03 &	1.39	$\pm$	0.05    &  1.159     & 78.81 (68) &	0.110\\
0.500	&	2.39 	$\pm$	0.02 	&  0.188 	$\pm$	0.003 	&	0.134 	$\pm$	0.03 &	1.42	$\pm$	0.04    &  1.185     & 80.56 (68) &	0.111\\
\hline
\hline
0.200	&	2.39 	$\pm$	0.02 	&  0.192 	$\pm$	0.003 	&	-0.834 	$\pm$	0.06 &	4.42 $\pm$	0.16    &  1.143     & 77.71 (68) &	0.258\\
0.322	&	2.39 	$\pm$	0.02 	&  0.191 	$\pm$	0.003 	&	-0.872	$\pm$	0.05 &	4.55	$\pm$	0.14    &  1.175     & 79.92 (68) &	0.263\\
0.400	&	2.40 	$\pm$	0.02 	&  0.192 	$\pm$	0.003 	&	-0.929	$\pm$	0.06 &	4.71	$\pm$	0.17    &  1.197     & 81.41 (68) &      0.267\\
0.500	&	2.40 	$\pm$	0.02 	&  0.191 	$\pm$	0.003 	&	-0.957	$\pm$	0.05 &	4.82 $\pm$	0.16    &  1.209     & 82.23 (68) &      0.271\\
\enddata
{Notes.\\ 
In columns 1-8, we show the following information: the hydrogen column density in units of $10^{22} \mathrm{cm}^{-2}$ ($N_{\mathrm{H}}$), the dimensionless photon index of power-law ($\Gamma$), the scattered fraction ($f_{\rm sc}$), the dimensionless spin parameter ($a_{*}$), the effective mass accretion rate of the disk in units of 10$^{18}$ g s $^{-1}$ ($\dot{M}$), the reduced chi-square ($\chi_{\nu}^{2}$), the total chi-square ($\chi^2$) and the degrees of freedom (d.o.f.), the bolometric Eddington-scaled luminosities ($L(a_{*}, \dot{M})/ L_{\mathrm{Edd}}$).\\
The two sets of system parameters: $M=5.7 M_{\odot}$, $i=70.0^{\circ}$, $D=6 \pm 2$~kpc (group 1) and $M=4.9M_{\odot}$, $i=80.0^{\circ}$, $D=6 \pm 2$~kpc (group 2) are separated with double horizontal lines. Above the double horizontal lines is group 1, below the double lines is group 2.}
\end{deluxetable*}

\section{4. DISCUSSION}
\label{section:4}

\begin{figure}[ht!]
\epsscale{0.7}
\centering
\plotone{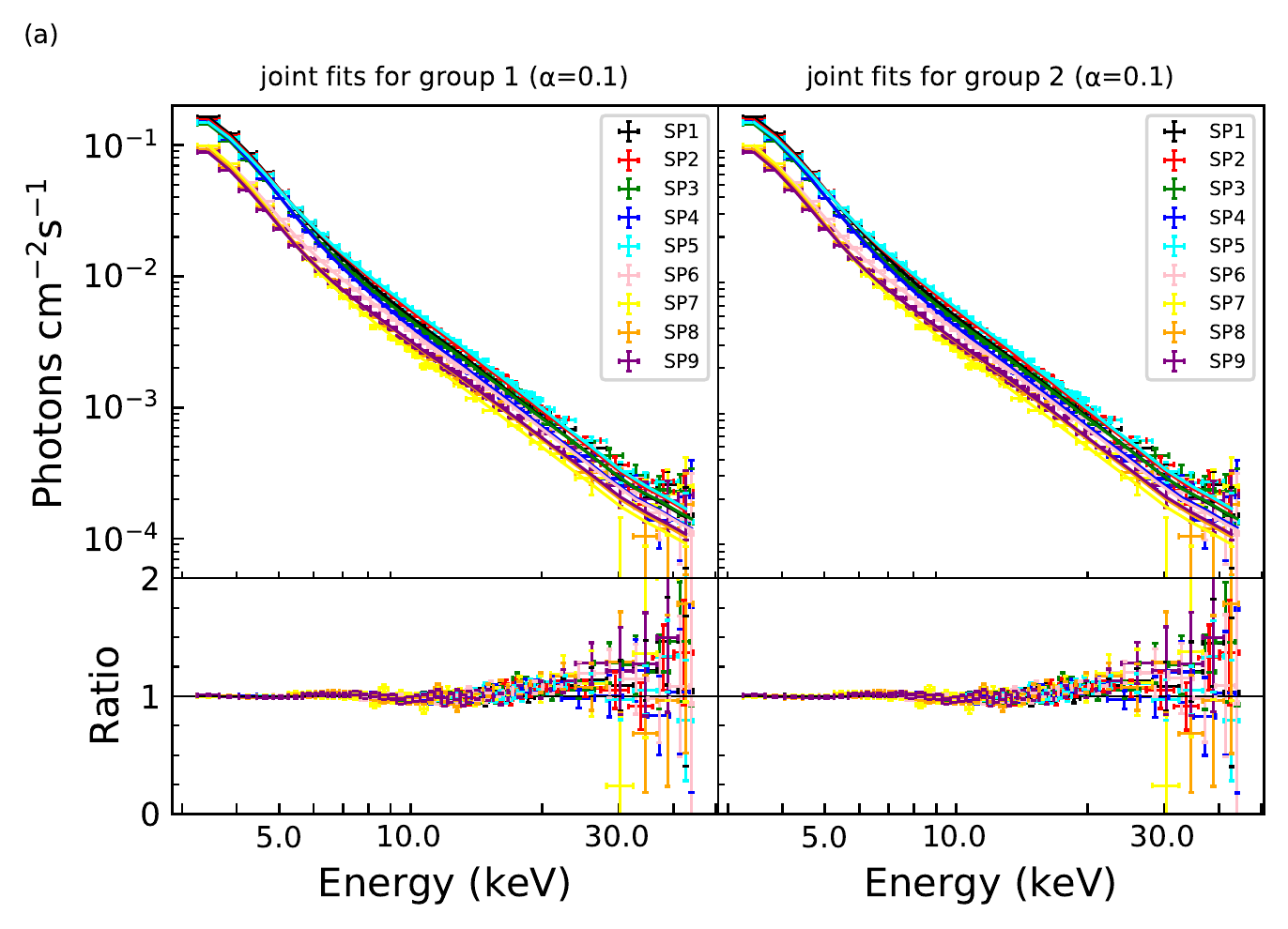}
\plotone{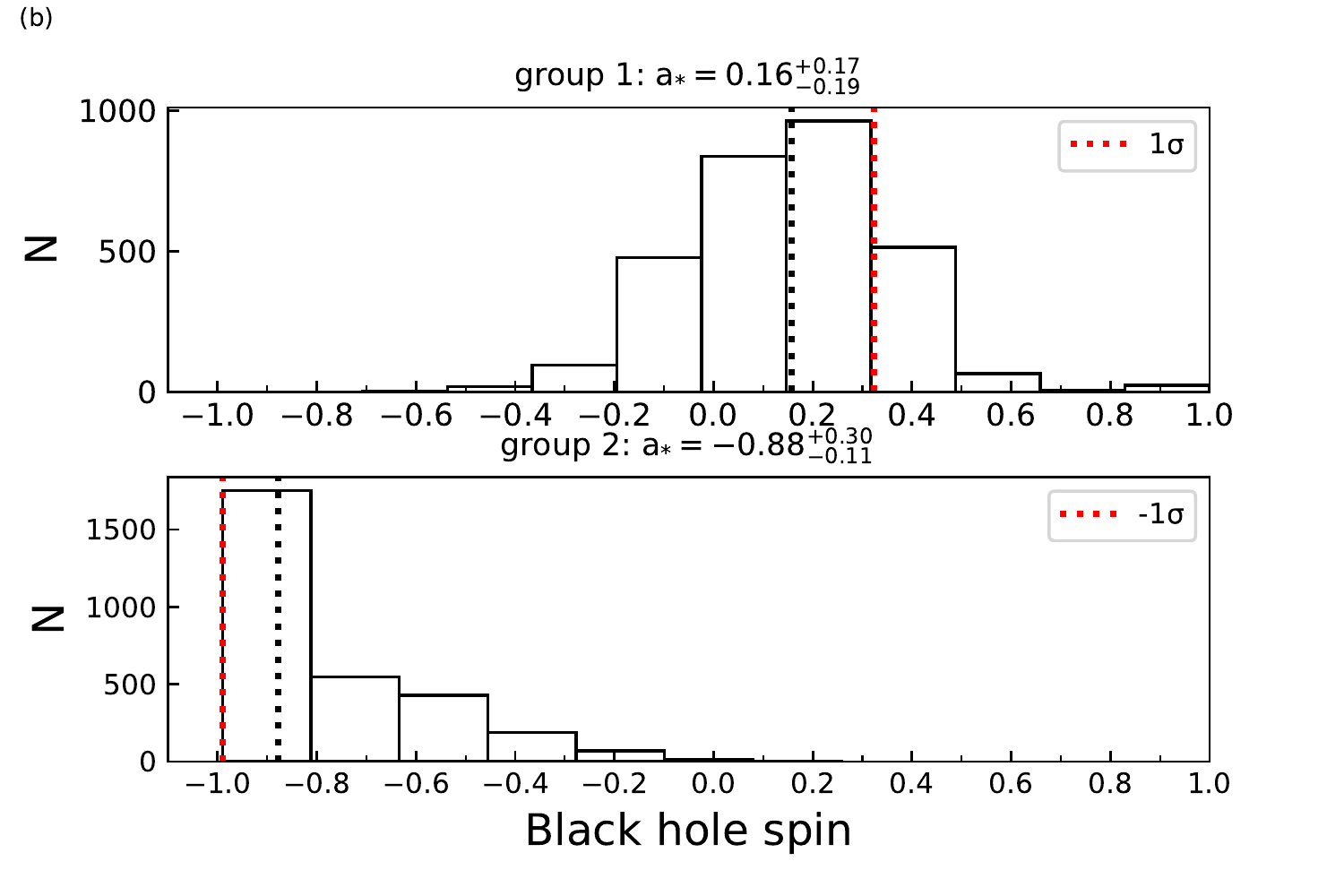}
\caption{(a) For $\alpha$=0.1, the linked spectral fit and its residuals over the fitted bandpass for relativistic model. Data have been rebinned for visual clarity. (b). Same as the Figure \ref{fig:5} definition, but as a MC result of joint fits.\label{fig:6}}
\end{figure} 

\subsection{4.1. Effect of the Hydrogen Column Density}
In order to assess whether the hydrogen column density significantly affects the spin results, we explore varying the value of $N_{\mathrm{H}}$ from 0.322 to 0.2, 0.4, 0.5 in units of $10^{22}\mathrm{cm}^{-2}$. These alternates are chosen as round numbers  covering the range of values given by \cite {Kennea2011}. The fitting is systematically affected for all the spectra in the same way. We illustrate this difference by showing such results for SP1 in Table \ref{table:5}. When $N_{\mathrm{H}}$ increases from $2\times10^{21}\mathrm{cm}^{-2}$ to $5\times10^{21}\mathrm{cm}^{-2}$, for group 1, $a_{*}$ varies from 0.210 to 0.134, with $\Delta a_{*}$ equaling to 0.076. And for group 2, $a_{*}$ changes from -0.834 to -0.957, and $\Delta a_{*}$ is 0.123. As expected, a slight increase of $N_{\mathrm{H}}$, causes the inferred spin decrease. Compared to the dynamical sources of uncertainty in hand, changing the value of $N_{\mathrm{H}}$ has very minor effect on the final spin results.

\begin{figure}[ht!]
\epsscale{0.7}
\centering
\plotone{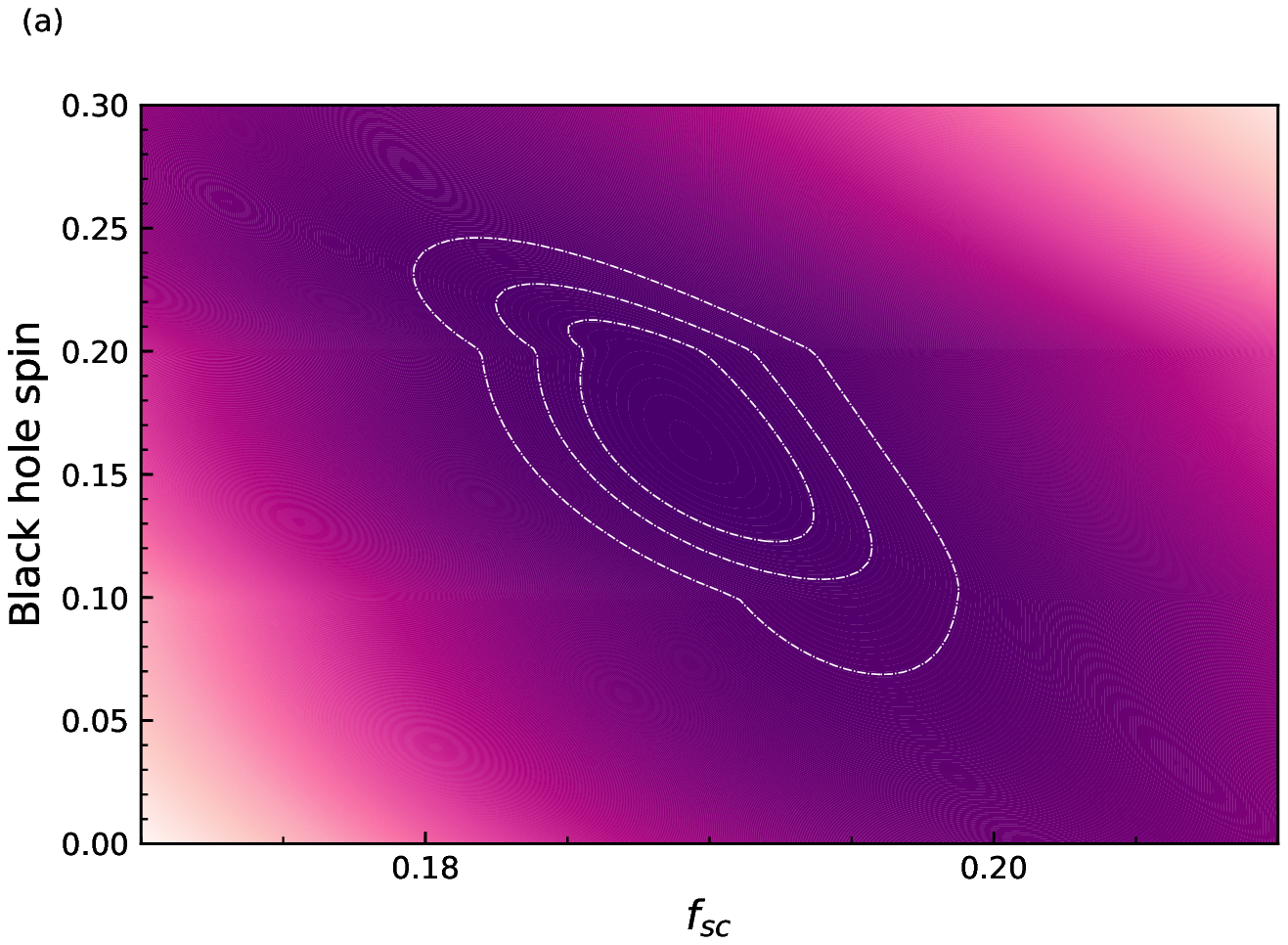}
\plotone{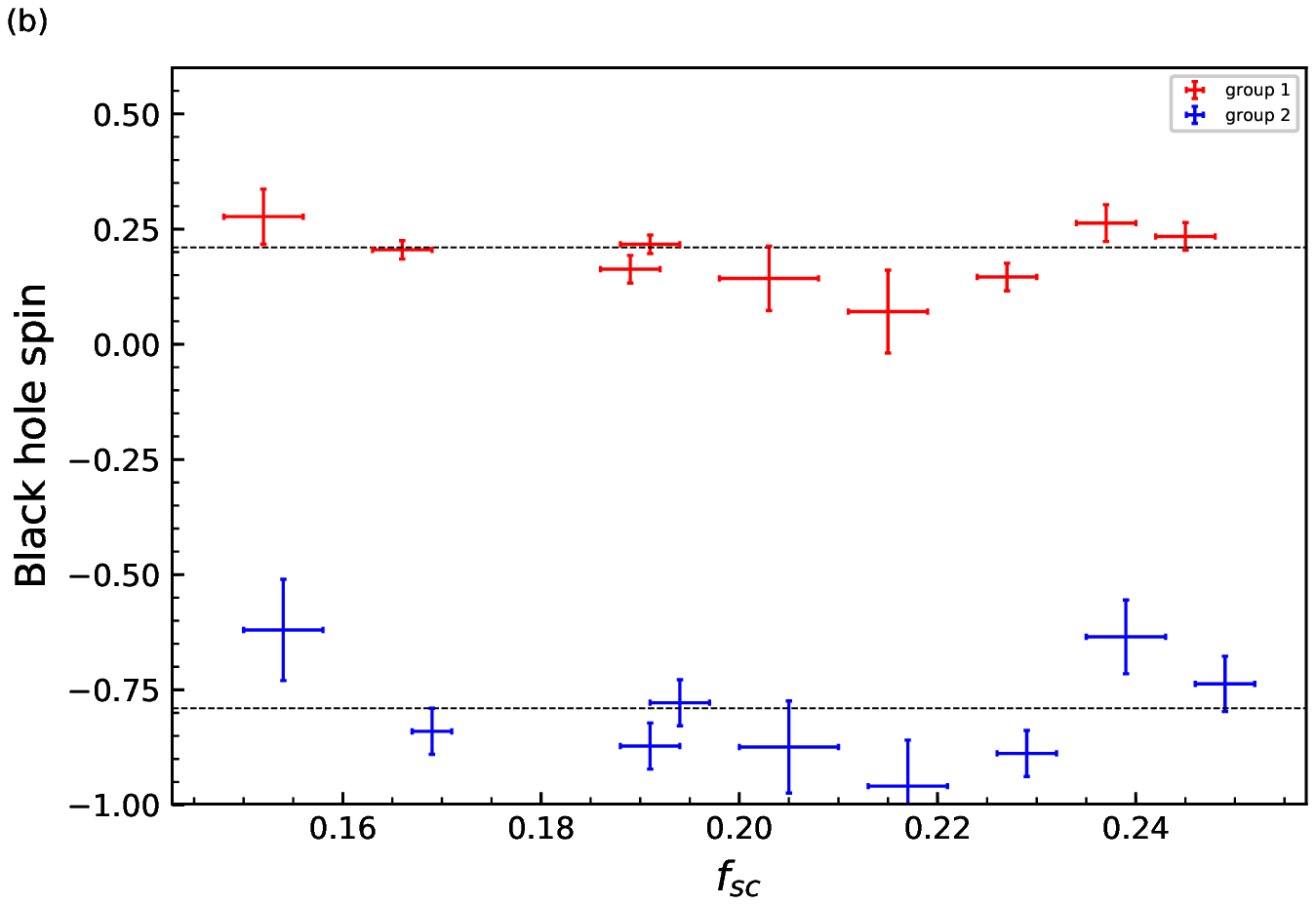}
\caption{(a). For SP1 under the default group 1 settings, the contour map of $f_{\mathrm{sc}}$ and $a_{*}$. The three contour lines represent $68.3\%$, 90 and $99.7\%$, respectively. (b). The relation between $a_{*}$ and $f_{\mathrm{sc}}$ for SP1 to SP9 in different groups.\label{fig:7}}
\end{figure}

\subsection{4.2. Effect of simultaneous fitting of 9 spectra on spin results}
In addition to fitting each spectrum individually with the relativistic model, we also consider the case that all the spectra are fitted simultaneously and looked into its effect. We linked the spin parameters among all 9 spectra, and letting the other fit parameters free. For the convenience of comparison, we also listed the results of simultaneous fits in Table \ref{table:3} and \ref{table:4} to group 1 and 2, and use a single horizontal line as a distinction. As a showcase, the simultaneous fit to all the spectra is also shown in Figure \ref{fig:6} (a). As we can see, the spin ranging from $-1 < a_{*} \lesssim 0.33$ (1$\sigma$) obtained from MC result (Figure \ref{fig:6} (b)), and it is clear that the results are fully consistent with ones obtained from individual spectral fits (Figure \ref{fig:5}), which is expected. Therefore, the effect of joint fits is negligible and our work uses the fit results from the individual spectra as our primary results.

\subsection{4.3. Differences from Previous Measurements}
\cite {Rout2020} construct a wider range of system parameter ($M$, $i$, $D$) grids based on an older set of measurement results. They found a simultaneous data set  with \emph{XMM}/EPIC-pn and \emph{RXTE}/PCA overlapping on September 28, 2010. Fitting the two data sets together, they simultaneously adopt the X-ray reflection fitting and continuum-fitting method. On this basis, they use meaningful values of mass accretion rate $\dot{M}$ to constrain the spin of MAXI J1659. In their results of the X-ray continuum-fitting show that the lower limit of the spin is pegged at -0.998, while the upper limit is 0.4. However, it should be noted that the spectra in \cite {Rout2020} do not meet the $f_{\mathrm{sc}} \lesssim 25 \%$ criterion. Different from \cite {Rout2020}, we firstly screen the spectra with $f_{\mathrm{sc}} \lesssim 25 \%$ for the X-ray continuum-fitting and made \texttt{crabcor} correction. In addition, we use the relativistic model \texttt{kerrbb2} that allows the spectral hardening factor $f$ to change, which will be closer to the actual situation. And we do not ignore the influence of self-irradiation in \texttt{kerrbb2}. Based on the use of different spectra, different models and updated dynamical parameters, we find a spin ranging from $-1 <  a_{*} \lesssim 0.35$ (1$\sigma$). As reported by \cite {Rout2020}, we also rule out extreme prograde spin.

\subsection{4.4. The possible relation between the scattering fraction in the spectrum and the spin}
To verify that a stronger hard tail does not introduce bias, we also check for possible correlation between the scattering fraction in the spectrum and the spin. We take SP1 as an example, and find for any individual spectrum, the contour map between black hole spin and $f_{\mathrm{sc}}$ shows a strong degenerate relation (see subgraph (a) of Figure \ref{fig:7}). However, we find that as an ensemble which encompasses a range of $f_{\mathrm{sc}}$ values, the spin does not exhibit correlation (see subgraph (b) of Figure \ref{fig:7}). Accordingly, we conclude that	$f_{\mathrm{sc}}$ does not affect our spin results.

\subsection{4.5. Further discussion on the Possibility of Extreme Negative Spin}
\cite {King1999} considered the effect of prograde accretion in changing the black hole mass and its spin (See their Figure 3). We also consider the influence of retrograde accretion in changing the black hole mass and spin in Appendix \ref{App:1}. As stated above, \cite {Kuulkers2013} found that the companion of MAXI J1659 had an initial mass of about $1.5~\mathrm{M}_{\odot}$, which evolved to its current mass in about 4.6-5.7 billion years. Let us make a simple estimate. We assume that MAXI J1659 is a retrograde black hole with an extreme spin, approaching $a_{*}=-1$. And we suppose the initial mass of the black hole is $4~\mathrm{M}_{\odot}$. The outburst-averaged dimensionless mass accretion rate ($\dot m=\dot M/\dot M_{\mathrm{Edd}}$) of the standard thin disk model is usually in the range of 0.01-0.3 (\citealt{Narayan1995}; \citealt{Ohsuga2002}), nonetheless, if we take into account that only one 7-month outburst has been observed from the source in the last 25 years, the time-averaged accretion rate $\dot m=2\times 10^{-4}$ would be more realistic. When $\dot m=2\times 10^{-4}$, the Eddington accretion rate ($\dot M_{\mathrm{Edd}}$) of a black hole with $4~\mathrm{M}_{\odot}$ is about $10^{-8}~\mathrm{M}_{\odot} ~\mathrm{yr}^{-1}$. According to Equation \ref{formula:1}, in 4.6-5.7 billion years, the accreted mass onto MAXI J1659 should fall within the range 0.0092-0.0114$~\mathrm{M}_{\odot}$ as $\dot m= $$2\times 10^{-4}$. In other words, $\Delta M/M_{\rm i}$ (defined in Appendix \ref{App:1}) should fall within the range of 0.0023-0.00285. This obviously is insufficient to rule out the possibility that MAXI J1659 has an extreme negative spin (see Figure \ref{fig:8}, two of the red dash lines represent 0.0023 and 0.00285, respectively). 

\section{5. CONCLUSION}
\label{section:5}
In this paper, we present an X-ray continuum-fitting spectral spectral analysis of the black hole candidate MAXI J1659. We select 9 spectra of \emph{RXTE}/PCA satisfying luminosity and state/coronal-brightness restrictions. Based on the two sets of plausible system parameters: $M=5.7 \pm 1.8~\mathrm{M}_{\odot}$, $i=70.0^{\circ}$, $D=6 \pm 2$ ~kpc and $M=4.9 \pm 1.6~\mathrm{M}_{\odot}$, $i=80.0^{\circ}$, $D=6 \pm 2$~kpc reported by \cite {Torres2021}, we constrain the spin to $-1 < a_{*} \lesssim 0.44$ in 90\% confidence interval via the X-ray continuum-fitting method, as suggested in \cite{Rout2020}, there is a possibility of an extreme negative spin. Then we exclude the influence of changing $N_\mathrm{H}$ on the spin. The possible effect of scattering fractions on spin results is also considered. We demonstrate that an extreme negative spin can not be ruled out on the basis of theoretically-expected accretion spin-up alone when considering the companion lifetime and a theoretical maximally retrograde black hole as a possible formation state. 

Besides, it is important to note that, taking the linked spectral for group 1 in $\alpha$=0.1 as an example, when we fix spin to its positive maximum value ($a_{*}=0.998$) and thaw inclination angle, we can obtain i=26.02 $\pm$ 0.17 degree, which is well below $i=70.0^{\circ}$ in group 1. This indicates that the inclination angle has a great impact on the spin. Therefore, accurate system parameters are very important for the CF method. It should be noted that, for MAXI J1659, more accurate spin results can be obtained when more precise measurements of system parameters are available in the future.

\acknowledgments
We thank the anonymous referee for the constructive comments. Y.F. thank the useful discussions with Prof. Tess Jaffe on extracting \emph{RXTE}/PCA spectrum. Y.F. also thank the useful discussions with Prof. Chang-Hwan Lee. The standard data products of \emph{RXTE}/PCA used in this work obtained from the \emph{RXTE} satellite and the \emph{RXTE} Guest Observer Facility (GOF). And the standard data products of \emph{XMM-Newton}/EPIC-pn obtained from the \emph{XMM-Newton} satellite, an ESA science mission funded by ESA Member States and the USA (NASA). The software used is provided by the High Energy Astrophysics Science Archive Research Centre (HEASARC), which is a service of the Astrophysics Science Division at NASA/GSFC and the High Energy Astrophysics Division of the Smithsonian Astrophysical Observatory. L.G. is supported by the National Program on Key Research and Development Project (Grant No. 2016YFA0400804), and by the National Natural Science Foundation of China (Grant No. U1838114), and by the Strategic Priority Research Program of the Chinese Academy of Sciences (Grant No. XDB23040100). J.W. acknowledges the support of the National Natural Science Foundation of China (NSFC grant No. U1938105) and the President Fund of Xiamen University (No. 20720190051).

\appendix
\section{retrograde accretion}\label{App:1}
Retrograde, meaning the black hole rotates in the opposite direction to its accretion disk ($a_{*}<0$). In order to keep the consistency with \cite {King1999}, $\Delta M$ is remained in use (where $\Delta M$ is the rest-mass added to the black hole from the initial state). And we assume the initial state of the black hole is $M=M_{\rm i}$, $a_{*}=-1$. Under the assumption of \cite {King1999}, for a retrograde black hole, through derivation, $\Delta M$ can be expressed as
\begin{equation}\Delta M=\left(\frac{{27M_{\rm i}}^2}{2}\right)^{1/2}\left[\sin^{-1}\left(\left(\frac {2M^2}{27M_{\rm i}^2}\right)^{1/2}\right)-\sin^{-1}\left(\left(\frac{2}{27}\right)^{1/2}\right)\right] \label{formula:1}\end{equation}
\begin{equation}a_{*}=\frac{M_{\rm i}}{M}\left[4-\left(\frac{27M_{\rm i}^{2}}{M^{2}}-2\right)^{1/2}\right]
 \label{formula:2} \end{equation}
When $a_{*}=0$, $M/M_{\rm i}=\left(3/2\right)^{1/2}$. The spin will increase as the accretion mass increases, indicated by formula \ref{formula:2}. And the speed of spin increasing is much faster than that of a prograde black hole. In other words, when the amount of spin changing is the same, the retrograde black hole accretion time is much shorter than the prograde black hole accretion time. 
 
\begin{figure}[ht!]
\epsscale{0.7}
\centering
\plotone{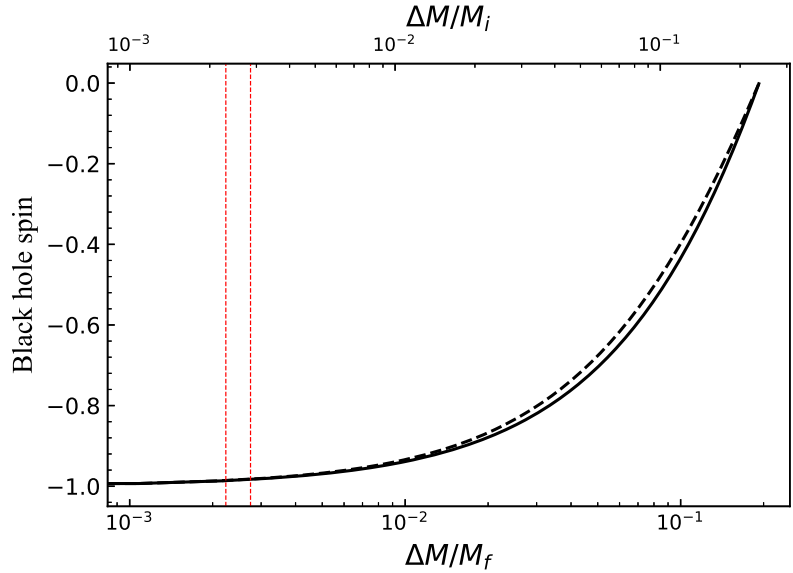}
\caption{$a_{*}$ versus accreted rest-mass $\Delta M$, in units of the final mass $M_{\rm f}$ at zero spin (solid line, bottom axis), and in units of the initial mass $M_{\rm i}\left(a_{*}=-1\right)$ at maximum negative spin (dotted line, top axis). Red dash lines represent $\Delta M/M_{\rm i}$=~0.0023 (left) and 0.00285 (right), respectively.\label{fig:8}}
\end{figure}

Combining formula \ref{formula:1} and \ref{formula:2}, the relation between $a_{*}$ and $\Delta M$ can be easily obtained, which shown in Figure \ref{fig:8}. When a maximum retrograde black hole is accreting the mass of its donor, and caused $a_{*}$ to change from -1 to 0. The mass that needs to accrete is $\Delta M\approx0.22M_{\rm i}$, where $M_{\rm i}$ represents the initial mass at $a_{*}=-1$. It can also be expressed as $\Delta M\approx0.18M_{\rm f}$, where $M_{\rm f}$ typifies the final mass at $a_{*}=0$.

\bibliography{ref}{}
\bibliographystyle{aasjournal}

\end{document}